\documentclass[aps,prb,twocolumn,floatfix,superscriptaddress,showpacs]{revtex4-1}
\usepackage{times}
\usepackage{epsfig}
\usepackage{amsmath}
\usepackage{amssymb}
\usepackage{graphicx}
\usepackage{bbold}
\usepackage{hyperref}
\hypersetup{colorlinks=true,citecolor=red,linkcolor=blue}

\newcommand{\tsh}{t_\mathrm{h}}

\newcommand{\bfk}{\mathbf{k}}
\newcommand{\bfa}{\mathbf{a}}

\newcommand{\bfr}{\mathbf{r}}
\newcommand{\bfA}{\mathbf{A}}
\newcommand{\bfR}{\mathbf{R}}
\newcommand{\bfB}{\mathbf{B}}

\newcommand{\udots}{\mathinner{\mskip1mu\raise1pt\vbox{\kern7pt\hbox{.}}
        \mskip2mu\raise4pt\hbox{.}\mskip2mu\raise7pt\hbox{.}\mskip1mu}}

\begin{document}
\title{Floquet engineering the Hofstadter butterfly in the square lattice and its effective Hamiltonian}
\author{Ming Zhao}
\affiliation{College of Physical Science and Technology, Guangxi Normal University, Guilin, Guangxi 541004}
\author{Qi Chen}
\affiliation{College of Physical Science and Technology, Guangxi Normal University, Guilin, Guangxi 541004}
\author{Xue-Dong Tian}
\affiliation{College of Physical Science and Technology, Guangxi Normal University, Guilin, Guangxi 541004}
\author{Liang Du}
\affiliation{College of Physical Science and Technology, Guangxi Normal University, Guilin, Guangxi 541004}
\affiliation{Department of Physics, Jiangxi Science and Technology Normal University, Nanchang, Jiangxi 330013}

\begin{abstract}
In this paper, we use Floquet theory to theoretically study the effect of
monochromatic circularly and linearly polarized light on the Hofstadter butterfly
in the square lattice, which is induced by uniform perpendicular magnetic field. In the absence of laser, the butterfly has a fractal, self-similar structure particle-hole symmetry and reflection symmetry about magnetic flux $\phi = 1/2$. These symmetries are preserved by the sub-lattice and the time-reversal symmetry, respectively. As the system is exposed to circularly polarized light, the original Hofsatdter butterfly in equilibrium is deformed by breaking both the particle-hole symmetry and the mirror symmetry, while the inversion symmetry about energy $E=0$ and magnetic flux $\phi=1/2$ is preserved. Our study show that, the circularly polarized light break both the sub-lattice symmetry and the time-reversal symmetry. The inversion symmetry is preserved because the Hamiltonian at magnetic flux $\phi$ and $1-\phi$ is connected through the sub-lattice transformation. Focusing on the small flux region, we study the Landau level and the influence of circularly polarized light on the Landau level. By considering the Floquet-Magus
expansion at high frequency, an quantitative description on the Landau level is given using numerical technique, which is determined by the one-photon absorption and emission process.
On the contrary, the linearly polarized light deforms the original Hofstadter butterfly by breaking the
rotational symmetry while preserving sub-lattice and the time-reversal symmetry.
Further, we study the influence of the periodic drive on the Chern number of the lowest band in middle Floquet copy within the off-resonance regime. We found strong circularly polarized light will change the Chern number. For linearly polarized light, the Chern number will not change and the values stay independent of laser polarization direction.
Our work highlights the generic features expected for the periodically driven Hofstadter problem on square lattice and provide the strategy to engineering the Hofstadter butterfly with laser.
\end{abstract}
\date{\today}
\maketitle

\section{INTRODUCTION}
\label{sec:intro}
The Hofstadter butterfly-a butterfly-like fractal energy diagram  theoretically predicted in hopping models on two dimensional square lattice subjected to a perpendicular uniform magnetic field-have been attracting great interest from the physics community\cite{Hofstadter:prb76}. The Hofstadter butterfly show particle-hole symmetry (reflection about zero energy axis) and reflection symmetry about magnetic flux $1/2$ (in unit of the fundamental flux quantum $hc/e$, where $h$ is Planck¡¯s constant, $c$ is the speed of light, and $e$ is the charge of the electron).
Physically, The reflection about zero energy axis is preserved by the bipartite symmetry of square lattice when only isotropic nearest-neighbor hopping is considered. The reflection symmetry about magnetic flux $1/2$ is preserved by the time-reversal symmetry (the lattice geometry do not distinguish $\pm z$ direction).
Theoretical study of the Hofstadter butterfly at the non-interacting level are its fractal structure dependence on Bravais lattices\cite{Hasegawa:prb06,LiJ:jpcm11,Hasegawa:prb13,Yllmaz:pra17}
(for examples, triangular,  rectangle, Honeycomb, Kagome lattice, etc.), non-uniform perpendicular magnetic field\cite{Oh:jkps00,LiJ:jpcm11}, three dimensional lattice\cite{Koshino:prl01} and disorder effect on Hofstadter butterfly\cite{ZhouC:prb05}.
Effect of electron-electron Coulomb interaction will modify the Hofstadter butterfly by changing its energy gap and bandwidth\cite{Gudmundsson:prb95,Doh:prb98,Kimura:prb02,Apalkov:prl14,LiangP:njp18}, not the characterized self-similar fractal structure.
Experimental observation of Hofstadter butterfly depend on the interplay of the magnetic field and the
lattice spacing. For the lattice spacing in typical atomic lattice (less than one nanometre), the required magnetic field strength to observe the butterfly is on the order of $10^4$ Tesla which is beyond current capabilities (about $10^2$ Tesla).
Artificial lattice (lattice spacing 100 nanometre) will require a small magnetic field, but it is too small to overcome the effect of disorder. A moir\'{e} super-lattices - created by put bilayer graphene coupled to hexagonal boron nitride (lattice spacing $10\mathrm{nm}$)
- is proposed to be a good candidate to observe the fractal structure in Hofstadter butterfly\cite{Dean:nat13, Hunt:sci13, Ponomarenko:nat13, Ni:cp2019}.

Recently, an interesting direction is to engineer the electronic band structure by dressing the bare electron with a periodic drive (photon absorption and emission). At non-interacting level, the system can be manipulated from non-topological phase to a topological one\cite{WangY:sci13,Kitagawa:prb10,Katan:prl13,Lindner:prb13,Fregoso:prb13,Dehghani:prb15a}. As the Coulomb interaction is taken into account, the effect of periodic drive can dress the electron by changing its strength of effective Coulomb interaction\cite{Tsuji:prb12, Gorg:nat18} (even inverse its sign).


At this point, the influence of periodic drive (lase in solid or temporal modulation in optical lattices) on the Hofstadter butterfly has not been studied extensively. \cite{WangJ:pra08,WangJ:jmo09,Lawton:jmp09,WangH:pre13,Lababidi:prl14, ZhouZ:prb14, Kooi:prb18, Wackerl:prb19} Pioneering study on the effect of periodic drive on the Hofstadter butterfly are mainly based on the kicked-Harper model\cite{WangJ:pra08,WangJ:jmo09,Lawton:jmp09}.
Prior work based on the square lattice Hofstadter model found that periodic driving (time-periodically modulate hopping integral in optical lattice) leads to pairs of counter-propagating chiral edge modes, which are protected by the chiral (sub-lattice) symmetry and robust against static disorder\cite{Lababidi:prl14, ZhouZ:prb14}.
In Ref.[\onlinecite{Wackerl:prb19}], the driven Hofstadter butterfly on the honeycomb lattice was studied by illuminating with monochromatic light (either circularly or linearly polarized light). Their study show that the influence of monochromatic light on the fractal structure of the spectrum depends mainly on its intensity and polarization.
The Chern numbers and the $W3$ invariants are numerically calculated to study the topology of the studied driven system.
Further, a more detailed study of the effect of a circularly polarized laser on the Hofstadter butterfly in honeycomb lattice is studied for different driving frequency regime\cite{Kooi:prb18}. In the Landau-level regime, new winglike gaps emerge as the driving frequency is decreased from off-resonance to resonant regime (laser frequency is smaller than bandwidth). Using a combination of spin wave theory and quantum field theory, a magnonic Floquet Hofstadter butterfly is realized in the two-dimensional (2D) insulating honeycomb ferromagnet\cite{Owerre:ap18}.
By considering different Bravais lattice, the driven Hofstadter in Kagome and triangular lattice are studied systematically\cite{du:prb18c}.

In these works, the fractal structure in Hofstadter butterfly are mainly numerically studied. The discussions are focusing on the symmetry broken and recovery process and its influence on deforming the butterfly-like structure.
For example, as the system is exposed to circularly polarized light, the reflection symmetry about magnetic flux $1/2$ is broken because circularly polarized light breaks the time-reversal symmetry.
In contrast, the time-reversal symmetry is preserved by the linearly polarized light and as a result, the mirror symmetry is preserved as the system is exposed to linearly polarized light.
In general, an analytical explanation of these numerical results will depend on its effective Hamiltonian.
In the limit of high frequency (off-resonance regime, laser frequency is larger than bandwidth), a periodically driven system can often be described by an effective static Hamiltonian using Floquet-Magus expansion\cite{Bukov:ap15,Mikami:prb16,Vogl:prb20}, which involves commutators between different Floquet copy. For the system with uniform magnetic field, the dimension of each Floquet copy will be $q$ (integer number, usually of order $10^2$ or larger). As a result, the effective Hamiltonian is usually hard to derive analytically (commutators of $q \times q$ matrix).

In this paper, we focus our study on the influence of periodic drive on the Hofstadter butterfly
on the two dimensional square lattice.
Using Floquet-Magus expansion, we derive the effective Hamiltonian for the Hofstadter butterfly deformed by off-resonance circularly (linearly) polarized light for the full magnetic flux regime.
Focusing on the small flux region, we study influence of circularly polarized light on the Landau level.
A analytical expression is given to describe the effect of periodic drive on the Landau level, which is determined by the one-photon absorption and emission process.
In addition, the linearly polarized light deforms the original Hofstadter butterfly by breaking the
rotational symmetry, which effectively change the ratio of hopping integral along $x$ and $y$.

The paper is organized as follows. In Sec.\ref{sec:model},  the model Hamiltonian on the two dimensional square lattice exposed to a perpendicular magnetic field and monochromatic laser is proposed. The Floquet theory is introduced subsequently. Numerical results on the influence of periodic driving laser (circularly and linearly polarized) on the Hofstadter butterlfy is illustrated in Sec.\ref{sec:benchmark}.
The effect of periodic laser on the Landau level is studied systematically in Sec.\ref{sec:landau}, and the Chern number is calculated in Sec.\ref{sec:chern}. Finally, in Sec.\ref{sec:conclusion}, we summarize our main results and conclusions.

\section{Model and method - square lattice}
\label{sec:model}
\begin{figure}[t]
\includegraphics[width=1.0\linewidth]{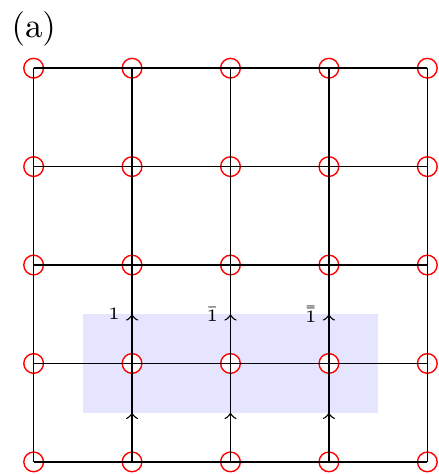}  %
\caption{(Color online) Square lattice with lattice constant $a$. The translational vectors are $\bfa_1=(1,0)a, \bfa_2=(0,1)a$ with one lattice point in unit cell. As the system is exposed to uniform perpendicular magnetic field, the unit cell is enlarged to recover translational symmetry. For magnetic flux $p/q$ (p,q are co-prime number), the enlarger one is $q$-times the original unit cell without magnetic field.}
\label{fig:square}
\end{figure}
The static model Hamiltonian is defined on the two dimensional square lattice with nearest-neighbour hopping terms (spin index are ignored for simplicity),
\begin{equation}
	H = \sum_{ij} \left( -t_x a^{\dagger}_{i,j} a_{i+1,j}^{} - t_y a^{\dagger}_{i,j} a_{i,j+1}^{}\right) + h.c.
\label{eq:Hsqr}
\end{equation}
where $t_x$ ($t_y$) is the hopping integral between nearest-neighbours along the $x$ ($y$) directions, $a^{\dagger}_{i,j} (a_{i,j}^{})$ create (annihilate) an electron at unit cell $\bfR(i,j)$.
The position of an arbitrary unit cell is
\begin{equation}
    \bfR(i,j) = i \bfa_{1} + j \bfa_{2}
\end{equation}
where $i,j$ are integers, $\bfa_1=(a,0)$ and $\bfa_2=(0,a)$ are the translational lattice vectors with $a$ the lattice constant. The electron dispersion can be derived by
Fourier transforming the Hamiltonian Eq.\eqref{eq:Hsqr} to momentum space,
\begin{equation}
	\epsilon_\bfk
           = -2t_x \cos(k_xa) - 2t_y \cos(k_y a)
\label{eq:epsilonk}
\end{equation}
As the system is exposed to field (electric, magnetic field) with a general vector potential $\bfA=(A_x, A_y)$,
the hopping terms in Eq.\eqref{eq:Hsqr} are modified through the Peierls substitution,
\begin{equation*}
    a_{i,j}^\dagger a_{i',j'}^{} \mapsto \exp\left(+i\frac{e}{\hbar c}\int_{{R}(i',j')}^{{R}(i,j)}
    \bfA\cdot d\bfr\right) a_{i,j}^\dagger a_{i',j'}^{}
\label{Peierls}
\end{equation*}
\subsection{Model Hamiltonian exposed to periodic drive (laser)}
As the system is exposed to periodic drive (laser) with vector potential $A_l(t) = [A_{l}^x(t), A_{l}^y(t)]$ (the subscript $l$ denote laser),
the time dependent Hamiltonian is written as,
\begin{equation}\begin{split}
	H(t) &= \sum_{ij} - t_x a^{\dagger}_{i,j} a_{i+1,j}^{}\exp(-i A_{l}^{x}(t)a) + h.c. \nonumber\\
         &+ \sum_{ij} - t_y a^{\dagger}_{i,j} a_{i,j+1}^{}\exp(-i A_{l}^{y}(t)a) + h.c.
\end{split}\end{equation}
Fourier transforming the Hamiltonian above to momentum space,
\begin{equation}\begin{split}
	H_\bfk(t) = - t_x \exp(ik_xa) \exp(-i A_l^x(t)a) a_{\bfk}^\dagger a_{\bfk} + h.c. \nonumber\\
                - t_y \exp(ik_ya) \exp(-i A_l^y(t)a) a_{\bfk}^\dagger a_{\bfk} + h.c.
\end{split}\end{equation}
Since the time-dependent Hamiltonian is time periodic $H_\bfk(t) = H_\bfk(t+T)$, where $T$ is the period of periodic drive.
According to the Floquet theory for time-periodic system, the wave function can be written as $|\Psi_\bfk(t)\rangle = \exp(-i\epsilon_{\bfk\alpha} t) |\phi_{\bfk\alpha}(t)\rangle$ with $|\phi_{\bfk\alpha}(t)\rangle = |\phi_{\bfk\alpha}(t+T)\rangle$. To calculate the quasi-energy $\epsilon_\alpha$, on can Fourier expand $|\phi_\bfk(t)\rangle = \sum_{m} \exp(im \Omega t) |\phi_{\bfk\alpha}^m\rangle$ with $|\phi_{\bfk\alpha}^m\rangle$ the $m$-th Floquet mode of Floquet wave function and $\Omega = 2\pi/T$.
The Schr\"{o}dinger equation is rewritten as,
\begin{equation}
    \sum_n (H_\bfk^{n,m} + m\delta_{m,n}\Omega)|\phi_\alpha^m\rangle = \epsilon_{\bfk\alpha} |\phi_\alpha^n \rangle
\end{equation}
where
\begin{equation}
	H_\bfk^{n,m} = H_\bfk^{n-m}= \frac{1}{T}\int_0^T dt H_\bfk(t) \exp(-i(n-m)\Omega t).
\end{equation}

For circularly polarized light with vector potential $A_{l}(t) = A_0 (\sin\Omega t, \cos\Omega t)$, we have
\begin{equation}\begin{split}
	&H_{\bfk}^{n-m}
                 = -t_x \exp(+ik_xa)\mathcal{J}_{m-n}(+A_0 a)a_{\bfk}^\dagger a_{\bfk}^{} \\
                  & -t_y \exp(+ik_ya)\mathcal{J}_{m-n}(+A_0 a)\exp(i(n-m)\pi/2) a_{\bfk}^\dagger a_{\bfk}^{}\\
                  & -t_x \exp(-ik_xa)\mathcal{J}_{m-n}(-A_0 a)a_{\bfk}^\dagger a_{\bfk}^{} \\
                  & -t_y \exp(-ik_ya)\mathcal{J}_{m-n}(-A_0 a)\exp(i(n-m)\pi/2) a_{\bfk}^\dagger a_{\bfk}^{}
\end{split}\end{equation}
with $\mathcal{J}_n$ the $n$-th order Bessel function of the first kind.
The effective Hamiltonian in the high frequency limit (Floquet-Magus expansion) reads,
\begin{equation}\begin{split}
	H_{\mathrm{eff}} \approx H_{0} + \frac{1}{\hbar\Omega} \sum_{l=1}^\infty[H_{l}, H_{-l}]_{-} + \mathcal{O}(\frac{1}{\hbar^2\Omega^2})
\end{split}\end{equation}
Due to the commutation relation $[H_{n}, H_{m}]_- = 0$, we have,
\begin{equation}\begin{split}
	H_{\mathrm{eff}} \approx H_{0} = \mathcal{J}_{0}(A_0 a)(-2t_x \cos(k_xa)-2t_y\cos(k_ya))a_{\bfk}^\dagger a_{\bfk} \nonumber
\end{split}\end{equation}
which is the original time independent Hamiltonian Eq.\eqref{eq:epsilonk} scaled by $\mathcal{J}_{0}(+A_0 a)$.

For Linearly polarized light with vector potential $A_{l}(t) = A_0 \sin\Omega t(\cos\alpha, \sin\alpha)$, we have
\begin{equation}\begin{split}
	H_{\bfk}^{nm}=
                  & -t_x \exp(+ik_xa)\mathcal{J}_{m-n}(+A_0 a\cos\alpha) a_{\bfk}^\dagger a_{\bfk} \nonumber\\
                  & -t_y \exp(+ik_ya)\mathcal{J}_{m-n}(+A_0 a\sin\alpha) a_{\bfk}^\dagger a_{\bfk}\nonumber\\
                  & -t_x \exp(-ik_xa)\mathcal{J}_{m-n}(-A_0 a\cos\alpha) a_{\bfk}^\dagger a_{\bfk} \nonumber\\
                  & -t_y \exp(-ik_ya)\mathcal{J}_{m-n}(-A_0 a\sin\alpha) a_{\bfk}^\dagger a_{\bfk}
\end{split}\end{equation}
The effective Hamiltonian in the high frequency limit (Floquet-Magus expansion),
\begin{equation}\begin{split}
	H_{\mathrm{eff}} \approx H_{0}
&= \mathcal{J}_{0}(A_0 a\cos\alpha)(-2t_x \cos(k_xa))a_{\bfk}^\dagger a_{\bfk} \nonumber\\
& + \mathcal{J}_{0}(A_0 a\sin\alpha)(-2t_y \cos(k_ya))a_{\bfk}^\dagger a_{\bfk}
\end{split}\end{equation}
which is the original time independent Hamiltonian Eq.\eqref{eq:epsilonk} with a scaling factor $\mathcal{J}_{0}(+A_0 a\cos\alpha)$ in the $x$-direction and $\mathcal{J}_{0}(+A_0 a\sin\alpha)$ in the $y$-direction, respectively. As we have $\alpha=\pi/4$, the effective Hamiltonian will be simplified by $\mathcal{J}_{0}(+A_0 a\cos\alpha)$ in both $x$- and $y$-direction.

\subsection{Hamiltonian subjected to perpendicular magnetic field}
As the system is subjected to a perpendicular magnetic field $\bfB =\nabla\times \bfA_m= (0,0, B)$ and
adopt the Landau gauge with vector potential $\bfA_m = (0, Bx, 0)$ (the subscript means magnetic field).
As usual, we restrict the  flux per
unit cell in units of the elementary charge over Planck's constant to a
rational value
\begin{equation}
    \phi \equiv B a^2.
\end{equation}
Thus, the Peierls phase can be written as
\begin{equation}
    \frac{e}{\hbar c} B a^2 = 2\pi \phi/\phi_0
\label{Peierls}
\end{equation}
where $\phi_0 = hc/e$ is the magnetic quantum flux, $\phi=p/q$ with $(p,q)$ co-prime integers.
To recover the translational symmetry of the lattice, we enlarge the unit cell along the translational vector $\bfa_1$ by $q$ times of the original unit cell without magnetic field. The Hamiltonian in real-space can be rewritten as,
\begin{equation}\begin{split}
 H = & -t_x \sum_{mn}\sum_{l=1}^{q-1} a^{\dagger}_{(i,j),l} a_{(i,j),l+1}^{} + h.c.\nonumber\\
     & -t_x \sum_{mn}\sum_{l=q}^{q} a^{\dagger}_{(i,j),q} a_{(i+1,j),1}^{} + h.c.\nonumber\\
     & -t_y \sum_{mn}\sum_{l=1}^{q} a^{\dagger}_{(i,j),l} a_{(i,j+1),l}^{} e^{-i2\pi l\phi} + h.c.
\end{split}\end{equation}
where the position of magnetic unit cell is
\begin{equation}
\tilde{\bf{R}}(i,j) = i \bfa_{1} \times q + j\bfa_{2}
\end{equation}
After Fourier transformation,
\begin{equation}\begin{split}
	H_{\bfk}
	 =&-t_x \left(\sum_{l=1}^{q-1} a^{\dagger}_{\bfk,l} a_{\bfk,l+1}^{}
	  + a^{\dagger}_{\bfk,q} a_{\bfk,1}^{} e^{i \bfk \cdot q\bfa_1}\right) + h.c. \nonumber\\
	 &-t_y \sum_{l=1}^{q} a^{\dagger}_{\bfk,l} a_{\bfk,l}^{}2\cos( \bfk \cdot \bfa_2-2\pi l\phi)
\end{split}\end{equation}
For the case $\bfk=(0,0)$, we have
\begin{equation}\begin{split}
	H_{\mathrm{butterfly}}
	 =&-t_x \left(\sum_{l=1}^{q-1} a^{\dagger}_{l} a_{l+1}^{}
	  + a^{\dagger}_{q} a_{1}^{} \right) + h.c. \\
	 &-t_y \sum_{l=1}^{q} a^{\dagger}_{l} a_{l}^{}2\cos( -2\pi l\phi)
	 \label{eq:eqbtfy}
\end{split}\end{equation}
This is the original Hamiltonian to generate the Hofstadter butterfly on the square lattice\cite{Hofstadter:prb76}.

\subsection{Hamiltonian with both laser and magnetic field}
As the system subjected to a perpendicular magnetic field is exposed to
periodic driving laser,
the Hamiltonian in real-space is expressed as,
\begin{equation}\begin{split}
 H = & -t_x \sum_{mn}\sum_{l=1}^{q-1} a^{\dagger}_{(m,n),l} a_{(m,n),l+1}^{}e^{-i \bfA_l(t) \cdot \bfa_1} + h.c.\nonumber\\
     & -t_x \sum_{mn}\sum_{l=q}^{q} a^{\dagger}_{(m,n),q} a_{(m+1,n),1}^{}e^{-i \bfA_l(t) \cdot \bfa_1} + h.c.\nonumber\\
     & -t_y \sum_{mn}\sum_{l=1}^{q} a^{\dagger}_{(m,n),l} a_{(m,n+1),l}^{} e^{-i2\pi l\phi}e^{-i \bfA_l(t) \cdot \bfa_2} + h.c.
\end{split}\end{equation}
Fourier transformation,
\begin{equation}\begin{split}
H_{\bfk} =& -t_x \sum_{l=1}^{q-1} a^{\dagger}_{\bfk,l} a_{\bfk,l+1}^{} e^{-i \bfA_l(t) \cdot \bfa_1} + h.c. \nonumber\\
	  & -t_x a^{\dagger}_{\bfk,q} a_{\bfk,1}^{} e^{i \bfk \cdot q\bfa_1} e^{-i \bfA_l(t) \cdot \vec{a}_1} + h.c. \nonumber\\
	  & -t_y \sum_{l=1}^{q} a^{\dagger}_{\bfk,l} a_{\bfk,l}^{} e^{i \bfk \cdot \bfa_2} e^{-i2\pi l\phi} e^{-i \bfA_l(t) \cdot \bfa_2} + h.c.
\end{split}\end{equation}
the Floquet-Bloch Hamiltonian in matrix form is expressed as,
\begin{equation}
H_{\mathrm{F}}(\phi) =
\begin{pmatrix}
    \ddots & \vdots & \vdots & \vdots & \udots \\
	\cdots & H_\bfk^0 - \hbar\Omega \mathbb{1} & H_\bfk^{-1} & H_\bfk^{-2} & \cdots \\
    \cdots & H_\bfk^1 & H_\bfk^0 & H_\bfk^{-1} & \cdots \\
    \cdots & H_\bfk^2 & H_\bfk^1 & H_\bfk^0 + \hbar\Omega \mathbb{1} & \cdots \\
    \udots & \vdots & \vdots & \vdots & \ddots
\end{pmatrix}
\label{HFml}
\end{equation}
with $\mathbb{1}$ the $q\times q$ unit matrix and
\begin{equation}
     H_\bfk^{nm} =  H_\bfk^{n-m} = \frac{1}{T} \int_0^T dt e^{-i(n-m)\Omega t}  H_\bfk(t).
     \label{eq:hfcopy}
\end{equation}
\subsubsection{Circularly polarized light}
For circularly polarized light with vector potential $\bfA(t)=A_0[\sin(\Omega t), \cos(\Omega t)]$,
The Floquet-Bloch Hamiltonian for circularly polarized light is expressed as,
\begin{equation}\begin{split}
	&H_{\bfk}^{nm}
	=  -t_x \mathcal{J}_{m-n}(A_0 a)\left(\sum_{l=1}^{q-1} a^{\dagger}_{\bfk,l} a_{\bfk,l+1}^{}
        + a^{\dagger}_{\bfk,q} a_{\bfk,1}^{} e^{i k_x  q a}\right)
         \nonumber\\
	  &-t_x \mathcal{J}_{m-n}(-A_0 a)\left(\sum_{l=1}^{q-1} a^{\dagger}_{\bfk,l+1} a_{\bfk,l}^{}
        +a^{\dagger}_{\bfk,q} a_{\bfk,1}^{} e^{i \bfk \cdot q\vec{a}_1} \right)
        \nonumber\\
	  & -t_y \mathcal{J}_{m-n}(+A_0a) \sum_{l=1}^{q} a^{\dagger}_{\bfk,l} a_{\bfk,l}^{}  e^{i k_y a} e^{-i 2\pi l\phi} e^{i(n-m)\pi/2} \\
	  &-t_y \mathcal{J}_{m-n}(-A_0a) \sum_{l=1}^{q} a^{\dagger}_{\bfk,l} a_{\bfk,l}^{} e^{-i k_y a} e^{i 2\pi l\phi} e^{i(n-m)\pi/2}
\end{split}\end{equation}

For the special case $\bfk=(0,0)$, The Hamiltonian is simplified as,
\begin{equation}\begin{split}
	H_{\bfk=0}^{nm}
	 = &-t_x \mathcal{J}_{m-n}(+A_0 a)\left( \sum_{l=1}^{q-1} a^{\dagger}_{l} a_{l+1}^{}
                                           + a^{\dagger}_{q} a_{1}^{} \right)  \\
      & -t_x \mathcal{J}_{m-n}(-A_0 a)\left(\sum_{l=1}^{q-1} a^{\dagger}_{l+1} a_{l}^{}
                                          + a^{\dagger}_{1} a_{q}^{} \right)  \\
	  & -t_y \mathcal{J}_{m-n}(A_0a)\sum_{l=1}^{q} a^{\dagger}_{l} a_{l}^{}   2\cos[2\pi l\phi-(n-m)\pi/2]
\label{eq:hfmaggcir}
\end{split}\end{equation}

For the high frequency limit, the effective Hamiltonian is expressed as
\begin{equation}\begin{split}
	H_{\mathrm{eff}} \approx H_{0} + \frac{1}{\hbar\Omega} [H_{1}, H_{-1}]_{-} + \frac{1}{\hbar\Omega} [H_{2}, H_{-2}]_{-}
\end{split}\end{equation}

\begin{figure*}[t]
\includegraphics[width=0.33\linewidth, angle=0]{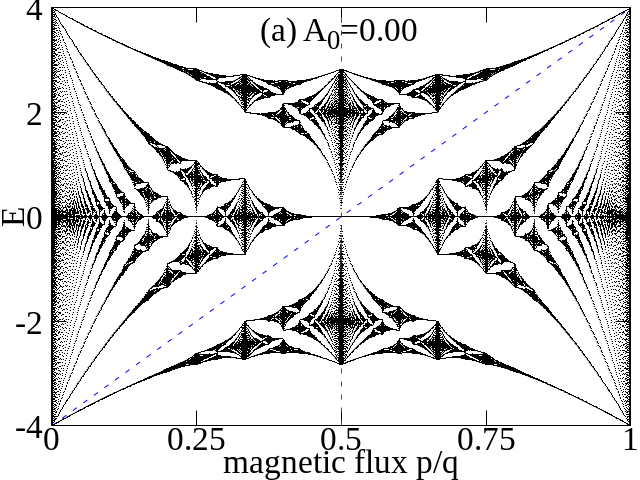}  %
\includegraphics[width=0.33\linewidth, angle=0]{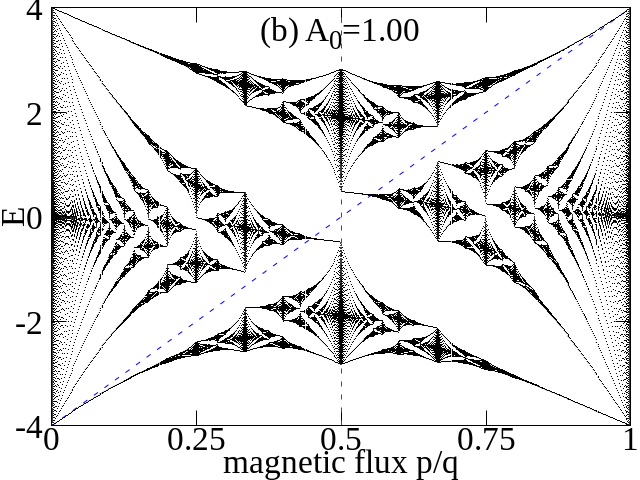}  %
\includegraphics[width=0.33\linewidth, angle=0]{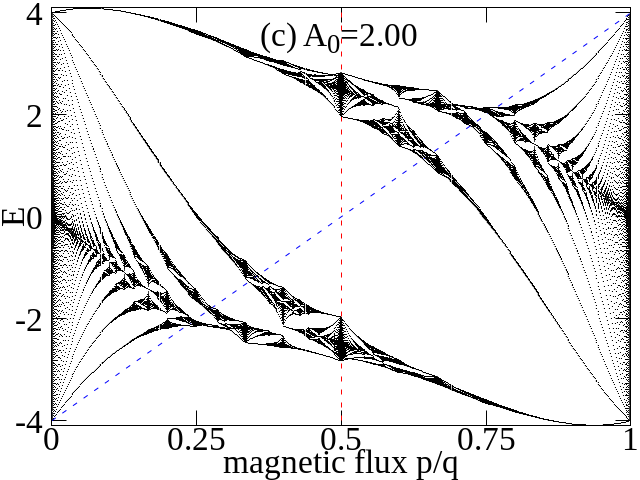}  %
\includegraphics[width=0.33\linewidth, angle=0]{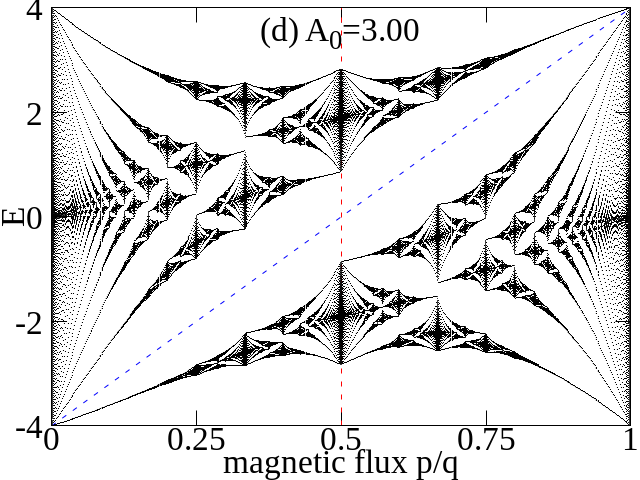}  %
\includegraphics[width=0.33\linewidth, angle=0]{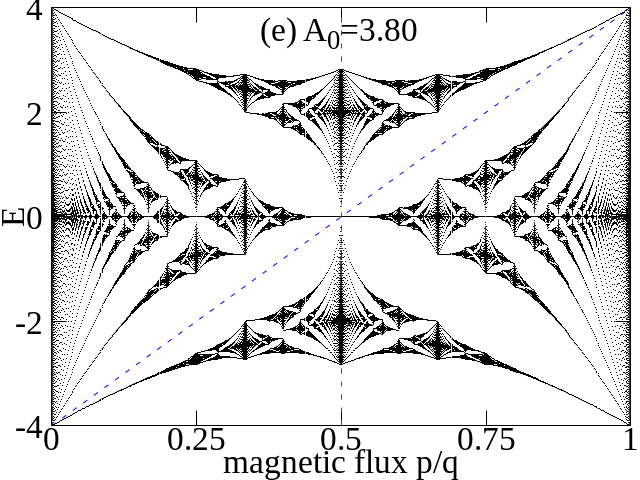}  %
\includegraphics[width=0.33\linewidth, angle=0]{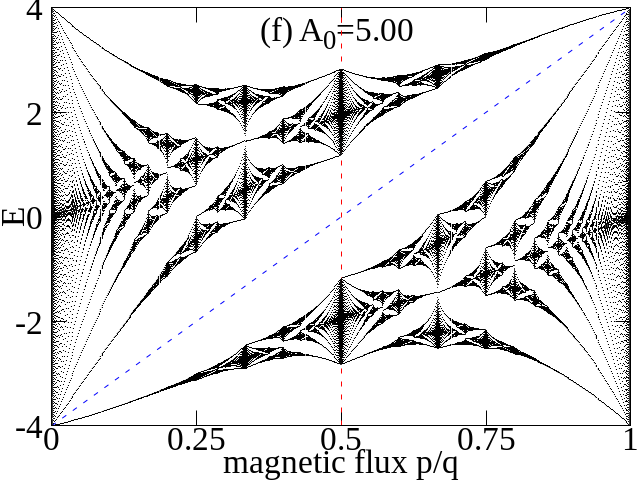}  %
\caption{(Color online.) The Hofstader butterfly for the square lattice, deformed by circularly polarized light  $\bfA(t) = A_0 (\sin\Omega t, \cos\Omega t)$ with frequency fixed at an off-resonant regime $\Omega = 8.0$. The representative laser intensity are chosen as (a) $A_0 = 0.0$, (b) $A_0 = 1.0$, (c) $A_0 = 2.0$ and (d) $A_0 = 3.0$, (e) $A_0 = 3.8$ and (f) $A_0 = 5.0$. The energy spectrum are re-scaled by $1/\mathcal{J}_0(A_0a)$ to have the same energy scale for different laser intensity. The calculations are done with $5$ Floquet copies. The magnetic flux is defined as $\phi = p/q$ with $p$ ranging from 1 to $q-1$ and $q=599$.
}
\label{fig:fig1}
\end{figure*}

\subsubsection{Linearly polarized light}
For the linear polarized light with vector potential $\bfA(t)=A_0\sin(\Omega t) (cos\alpha, sin\alpha)$,
the Floquet-Bloch Hamiltonian is written as,
\begin{widetext}
\begin{equation}\begin{split}
	&H_{\bfk}^{nm}
	 = -t_x \mathcal{J}_{m-n}(A_0 a\cos\alpha) \left(\sum_{l=1}^{q-1} a^{\dagger}_{\bfk,l} a_{\bfk,l+1}^{}  + a^{\dagger}_{\bfk,q} a_{\bfk,1}^{} e^{+i \bfk \cdot q\vec{a}_1}\right)  \nonumber\\
	  &  -t_x \mathcal{J}_{m-n}(-A_0 a\cos\alpha) \left(\sum_{l=1}^{q-1} a^{\dagger}_{\bfk,l+1} a_{\bfk,l}^{}  + a^{\dagger}_{\bfk,1} a_{\bfk,q}^{} e^{-i \bfk \cdot q\vec{a}_1}\right) \nonumber\\
	  & -t_y \sum_{l=1}^{q} a^{\dagger}_{\bfk,l} a_{\bfk,l}^{} \mathcal{J}_{m-n}(+A_0a\sin\alpha) \exp(+i (k_y a - 2\pi l\phi))\nonumber\\
      & -t_y \sum_{l=1}^{q} a^{\dagger}_{\bfk,l} a_{\bfk,l}^{} \mathcal{J}_{m-n}(-A_0a\sin\alpha) \exp(-i (k_y a - 2\pi l\phi))\nonumber\\
\end{split}\end{equation}
\end{widetext}
For the special case $\bfk=(0,0)$, The Hamiltonian is simplified as,
\begin{equation}\begin{split}
	&H_{\bfk=0}^{nm}
	 = -t_x \mathcal{J}_{m-n}(+A_0 a\cos\alpha) \left(\sum_{l=1}^{q-1} a^{\dagger}_{l} a_{l+1}^{} + a^{\dagger}_{q} a_{1}^{} \right) \\
	  &  -t_x \mathcal{J}_{m-n}(-A_0 a\cos\alpha) \left(\sum_{l=1}^{q-1} a^{\dagger}_{l+1} a_{l}^{}  + a^{\dagger}_{1} a_{q}^{} \right) \\
	  & -t_y \sum_{l=1}^{q} a^{\dagger}_{l} a_{l}^{} \mathcal{J}_{m-n}(+A_0a\sin\alpha) \exp(-i 2\pi l\phi)\\
      & -t_y \sum_{l=1}^{q} a^{\dagger}_{l} a_{l}^{} \mathcal{J}_{m-n}(-A_0a\sin\alpha) \exp(+i 2\pi l\phi)
      \label{eq:hflin}
\end{split}\end{equation}

\section{Hofstadter butterfly}
\label{sec:benchmark}
In the equilibrium case (absence of laser), the Hofstadter butterfly on the isotropic square lattice (only nearest-neighbor hopping) is calculated with $q=599$ (shown in Fig.\ref{fig:fig1} wih $A_0=0.00$).
The energy diagram show complex fractal structure with rich symmetries. Three of its rich symmetries will be discussed explicitly below.

Firstly, the reflection symmetry about zero energy axis (particle-hole symmetry) is observed. This symmetric properties is explained by considering the sub-lattice symmetry (square lattice is a bipartite lattice, the sub-lattice symmetry implies that energies must occur in pairs of $\pm E$), which is,
\begin{equation}
    S(H(\phi))  = -S(H(\phi)),
    \label{eq:subsym}
\end{equation}
where $S(H(\phi))$ means the spectrum of Hamiltonian $H(\phi)$.
Note here the exact particle-hole symmetry is observed only when $q$ is even or A-B sublattice is considered in one unit cell\cite{}.

Secondly, the reflection symmetry about magnetic flux $\phi=1/2$ is observed, which means $\epsilon_n(\phi) = \epsilon_n'(1-\phi)$ with $\epsilon_n(\phi)$ the $n$-th eigenvalue of the Hamiltonian Eq.\eqref{HFml} with magnetic flux $\phi$.
This symmetry can be explained using the time-reversal operator. Since magnetic field change sign under time-reversal operation. The time reversal partner of the Hamiltonian at magnetic flux $\phi$ is the one with magnetic flux $-\phi$, which is, $\mathcal{T} H(\phi)\mathcal{T}^{-1} =  H(-\phi)$ with $\mathcal{T}$ the time reversal operator (anti-unitary). In addition, we have $H(-\phi) = H(1-\phi)$ due to the periodicity of the magnetic flux\cite{Osadchy:jmp01,Yllmaz:pra17}.
Finally, we get,
\begin{equation}
    \mathcal{T} H(\phi) \mathcal{T}^{-1} = H(-\phi) = H(1-\phi)
    \label{eq:periodmag}
\end{equation}
Since the two operators that are time-reversal partners will have the same eigenvalues, we have the property for spectrum,
\begin{equation}
    S(H(\phi))  = S(H(-\phi)) = S(H(1-\phi)),
    \label{eq:trsym}
\end{equation}
which explain the reflection symmetry about magnetic flux $\phi=1/2$. Physically, we can understand the symmetry  $\epsilon_n(\phi)  = \epsilon_n(-\phi)$ by thinking that the lattice geometry do not distinguish $\pm z$ direction.

Thirdly, by considering both time-reversal Eq.\eqref{eq:trsym} and sub-lattice symmetry Eq.\eqref{eq:subsym}, we have,
\begin{equation}
    S(H(\phi))  = -S(H(1-\phi)),
    \label{eq:invsym}
\end{equation}
which imply the inversion symmetry about point $E=0,\phi=1/2$. A more detailed calculation (proof) is elaborated in Apppendices.
\subsection{Effect of an off-resonance circularly polarized light}
As the Hofstadter butterfly is further exposed to a circularly polarized light, the self-similar fractal structure is deformed.
Here several representative laser intensity are chosen as $A_0=1.0, 2.0, 3.0, 3.8, 5.0$ to exhibit the deformation of Hofstadter's butterfly by laser. The laser frequency is fixed as $\hbar\Omega = 8$. Note, in order to compare with the Hofstadter butterfly without laser, the spectrum with finite laser intensity is scaled by $1/\mathcal{J}(A_0a)$.

Qualitatively, let's focus on the above three symmetries (particle-hole, reflection about $\phi=1/2$ and inversion about $E=0,\phi=1/2$) in the original Hofstadter butterfly ($A_0 = 0.00$ in Fig.\ref{fig:fig1}).
By a quick look at the deformed butterfly with laser intensity $A_0 = 1.00$ in Fig.\ref{fig:fig1}, the particle-hole symmetry and the reflection symmetry about axis $\phi=1/2$ are broken while the inversion symmetry about $\phi=1/2, E=0$ is preserved and a physical explanation is as follows.
The particle-hole symmetry is absent because the effect of circularly polarized light periodic driving laser break the sub-lattice symmetry through the photon absorption and emission processes\cite{Oka:prb09}.
The reflection symmetry about axis $\phi=1/2$ no longer exist because the circularly polarized light break the time-reversal symmetry (time-reversal partner of left polarized light will be right polarized one). In other words, since the circularly polarized light will define the $z$-direction for the lattice, changing the direction of magnetic field $B$ ($\phi$ is changed to $-\phi$) will break the reflection symmetry about $\phi=1/2$ (periodicity of the magnetic
flux $H(-\phi) = H(1-\phi)$ is used).
Surprisingly, even though both the sub-lattice and the time reversal symmetries are broken, the inversion symmetry of the butterfly about the point $\phi=1/2, E=0$ is preserved.

By comparing the deformed Hofstadter butterfly with increasing laser intensity, we realize that the three symmetries in original butterfly can be recovered approximately at critical laser intensity $A_0 = 3.80$. Further increasing the intensity will break the first two symmetries while preserve the third symmetry again.
This kind of behavior can be understood by considering the effective Hamiltonian under high frequency expansion\cite{}.

\begin{figure*}[t]
\includegraphics[width=0.33\linewidth, angle=0]{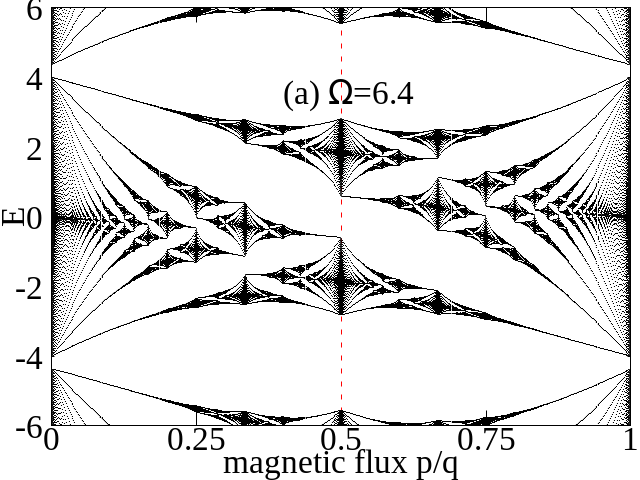}  %
\includegraphics[width=0.33\linewidth, angle=0]{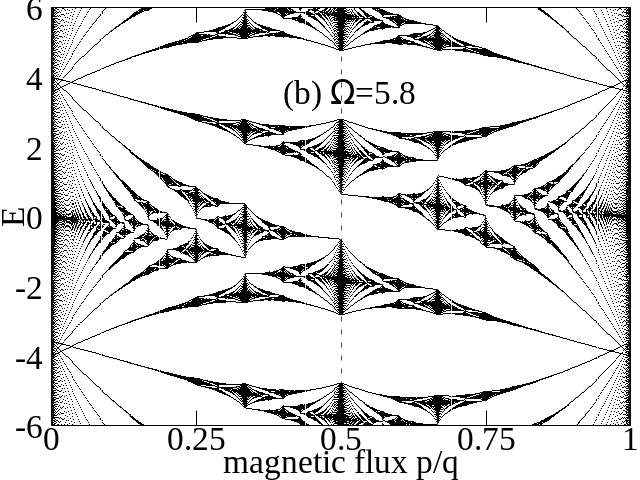}  %
\includegraphics[width=0.33\linewidth, angle=0]{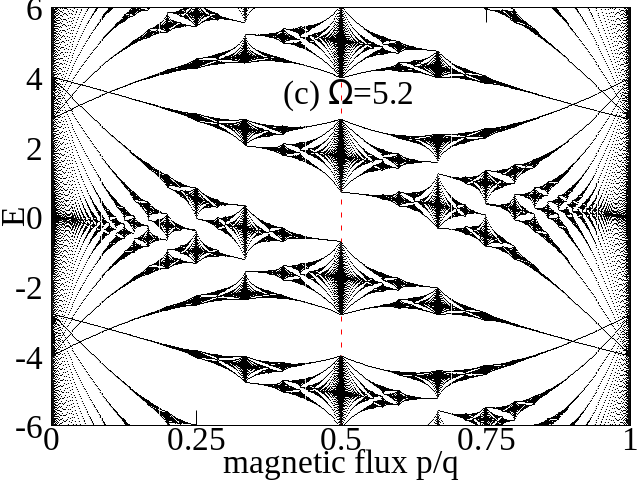}  %
\caption{(Color online.) The Hofstader butterfly for the square lattice, deformed by circularly polarized light  $\bfA(t) = A_0 (\sin\Omega t, \cos\Omega t)$ with laser intensity fixed at $A_0a = 1.0$. The laser frequency is varying from off-resonance to on-resonance regime (a) $\hbar\Omega = 6.4$, (b) $\hbar\Omega = 5.8$, (c) $\hbar\Omega = 5.2$. The energy spectrum are re-scaled by $1/\mathcal{J}_0(A_0a)$ to have the same energy scale for different laser intensity. The calculations are done with $5$ Floquet copies. The magnetic flux is defined as $\phi = p/q$ with $p$ ranging from 1 to $q-1$ and $q=599$.
}
\label{fig:onresonance}
\end{figure*}
\begin{figure*}[t]
\includegraphics[width=0.33\linewidth, angle=0]{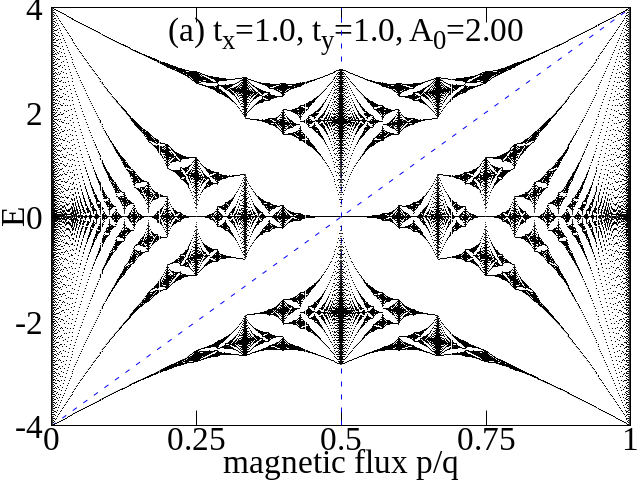}  %
\includegraphics[width=0.33\linewidth, angle=0]{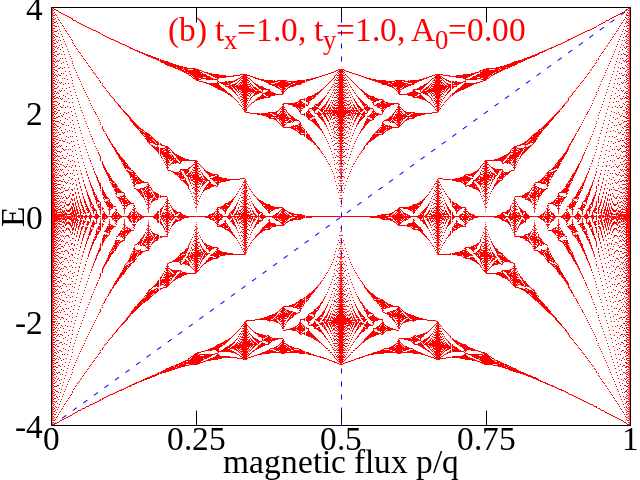}  %
\includegraphics[width=0.33\linewidth, angle=0]{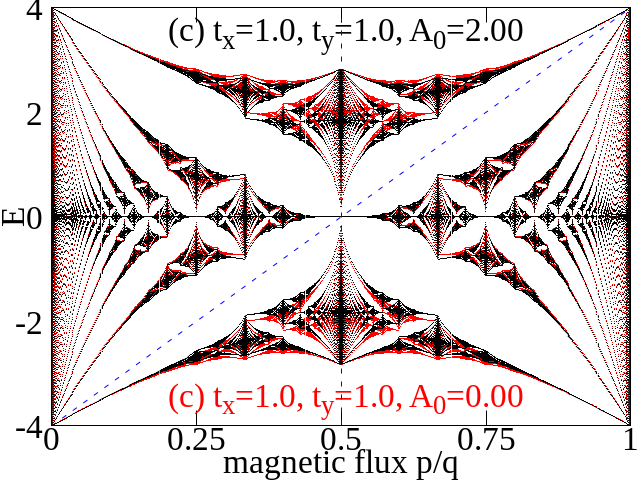}  %
\includegraphics[width=0.33\linewidth, angle=0]{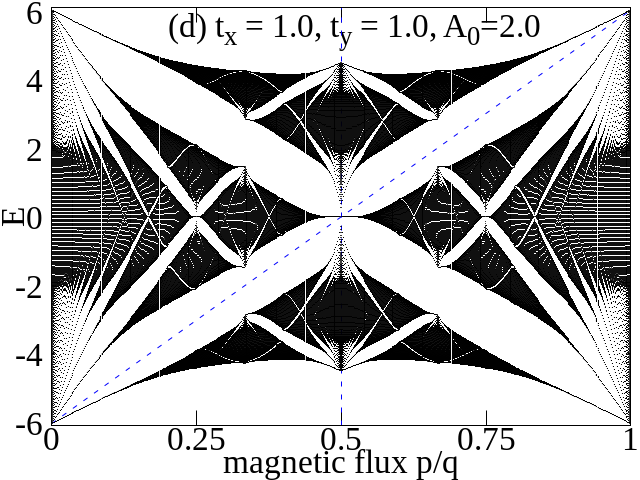}  %
\includegraphics[width=0.33\linewidth, angle=0]{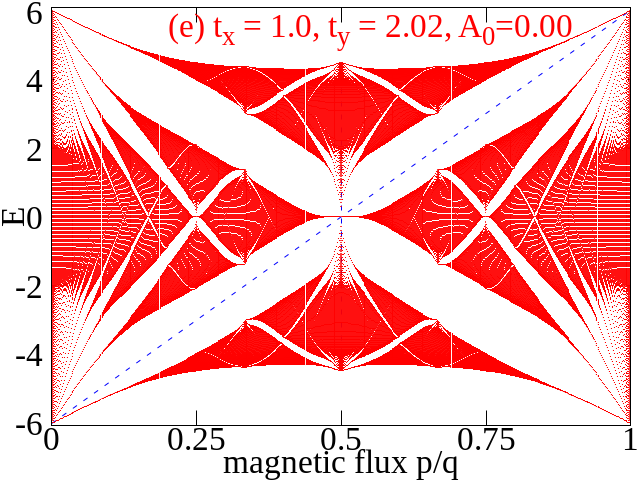}  %
\includegraphics[width=0.33\linewidth, angle=0]{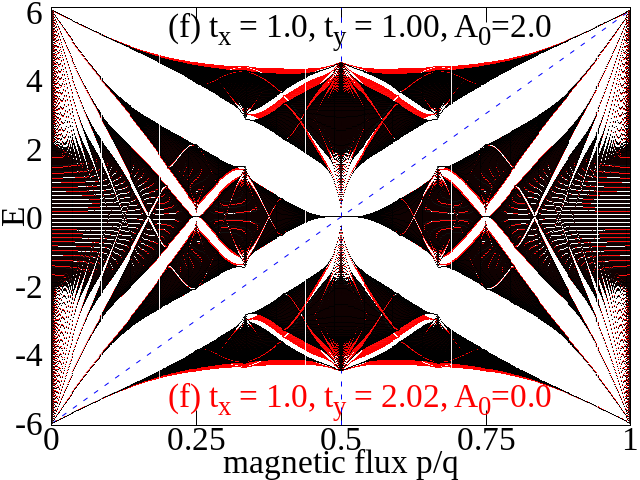}  %
\caption{(Color online.) The Hofstader butterfly for the square lattice, deformed by linearly polarized light  $\bfA(t) = A_0 \sin\Omega t (\cos\alpha, \sin\alpha)$ with frequency fixed at an off-resonant regime $\Omega = 8.0$. The representative laser intensity are chosen as $A_0 = 2.0$.
The energy spectrum are re-scaled by $1/\mathcal{J}_0(A_0a \cos\alpha)$ to have the same energy
scale for different laser intensity. The calculations are done with $5$ Floquet copies.
The magnetic flux is defined as $\phi = p/q$ with $p$ ranging from 1 to $q-1$ and $q=599$.
(a) $t_x =1.0, t_y =1.0,A_0=2.0, \alpha=\pi/4$, (b) butterfly at equilibrium case (absence of laser) $t_x =1.0, t_y =1.00,A_0=0.0$, (c) a comparison of (a) and (b);
(d) $t_x =1.0, t_y =1.0,A_0=2.0, \alpha=\pi/6$, (e) butterfly at equilibrium case (absence of laser) $t_x =1.0, t_y =2.02,A_0=0.0$, (f) a comparison of (d) and (e).
}
\label{fig:fig5}
\end{figure*}
In the high frequency limit, the effective Hamiltonian\cite{Bukov:ap15,Mikami:prb16,Vogl:prb20} is expressed as
\begin{equation}\begin{split}
	H_{\mathrm{eff}}(\phi) \approx& H_{0}(\phi) + H'(\phi) + \mathcal{O}(1/\hbar^2\Omega^2) \nonumber\\
\end{split}\end{equation}
where $H_{0}(\phi) = \mathcal{J}_0(A_0a)H_{\mathrm{butterfly}}$  and
\begin{widetext}
\begin{equation}\begin{split}
    H'(\phi) =  [H_{1}, H_{-1}]_{-} / (\hbar\Omega)
            =& 8 t_xt_y\mathcal{J}_{1}^2(A_0 a)\sin(\pi \phi)
	\left(\sum_{l=1}^{q-1}a^{\dagger}_{\bfk,l} a_{\bfk,l+1}^{}\cos(2\pi (l+1/2)\phi)
	+ a^{\dagger}_{\bfk,q} a_{\bfk,1}^{}\cos(+ \pi \phi)\right)\\
\end{split}\end{equation}
\end{widetext}
We have $H_{0}\propto \mathcal{J}_0(A_0a)$ and $H' \propto \mathcal{J}_1^2(A_0a)$. And as a result,
the interplay of $\mathcal{J}_1^2(A_0a)$ and $\mathcal{J}_0(A_0a)$ play an important role. At the zero point of $\mathcal{J}_0(A_0a)$, the first order correction will dominate the behavior of energy spectrum. While at the zero points of $\mathcal{J}_1(A_0a)$ where $A_0a=3.83$ ($\mathcal{J}_0(A_0a)$ remains finite), the effective Hamiltonian will be
the equilibrium one scaled by a Bessel function  $\mathcal{J}_0(A_0a)$, where both the reflection symmetry about $E=0$ axis and $\phi=1/2$ axis are recovered approximately. Further increasing the laser intensity above critical one will break the two symmetries again.
\subsection{Effect of an on-resonance circularly polarized light}
Follow the study in Ref.[\onlinecite{Kooi:prb18}], we consider influence of on-resonance circularly polarized laser on the Hofstadter butterfly in this subsection. Here we choose the laser intensity to be $A_0 = 1.0$ in Fig.\ref{fig:onresonance}. The effective bandwidth for the butterfly will be $8\mathcal{J}(1.0) =  6.1216$. So we choose the representative laser frequency to be around the effect bandwidth $\hbar\Omega = 6.40, 5.80, 5.20$. Here We observe the overlap between different Floqquet copy and the crossing of Landau levels at small flux regime. The avoided crossing is not observed here.

\subsection{Effect of an off-resonance linearly polarized light}
The Hofstadter butterfly deformed by linearly polarized light is plotted in Fig.\ref{fig:fig5}. Here we fix the laser intensity to be $A_0a=1.0$ and the frequency in the off-resonance regime $\hbar\Omega=8.0$. The vector potential of a linearly polarized laser is $\bfA(t) = A_0(\cos\alpha, \sin\alpha))$.
By considering the polarization direction, we plot the $\alpha=\pi/4$ in the upper panels in Fig.\ref{fig:fig5}.
Note here, the energy are rescaled by $1/\mathcal{J}(A_0a\cos\alpha)$ for the purpose to have the same energy scale with original Hofstadter without laser.
By comparing with the two data, We find they are very similar with each other, except main difference in the larger magnetic field (around $\phi=1/2$).
This feature can be understood by considering the effective Hamiltonian in high frequency regime.
From the expression in  Eq.\eqref{eq:hflin}, the Floquet Hamiltonian have the property,
\begin{equation}
     H_{+n} = (-1)^n H_{-n}
\end{equation}
As a result, the effective Hamiltonian at high frequency can be simplified as,
\begin{equation}\begin{split}
	H_{\mathrm{eff}} \approx& H_{0} + \frac{1}{\hbar\Omega} \sum_{n=1}^\infty[H_{n}, H_{-n}]_{-} + \mathcal{O}(\frac{1}{\hbar^2\Omega^2}) \\
	=& H_0 + \mathcal{O}(\frac{1}{\hbar^2\Omega^2})
\end{split}\end{equation}
where the hopping integral along $x$-direction is re-normalized by $\mathcal{J}(A_0a\cos\alpha)$, the $y$-direction re-normalized by $\mathcal{J}(A_0a\sin\alpha)$. To confirm our conclusion above, we plot the deformed Hofstadter butterfly with a different polarization direction $\alpha=\pi/6$.  The calculated result coincide with the data for $t_x = 1.00, t_y = 2.02$ which is $t_y/t_x = \mathcal{J}(A_0a\sin\pi/6)/\mathcal{J}(A_0a\cos\pi/6)$ for $A_0a=2.0$.
\begin{figure*}[t]
\includegraphics[width=0.245\linewidth, angle=0]{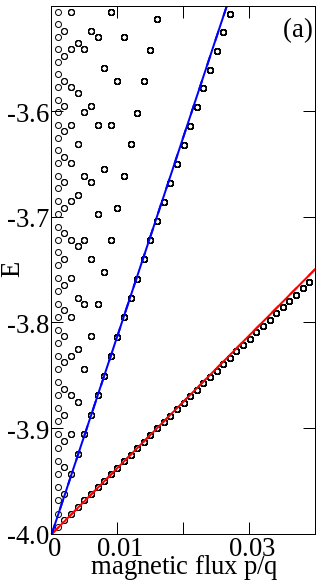}  %
\includegraphics[width=0.740\linewidth, angle=0]{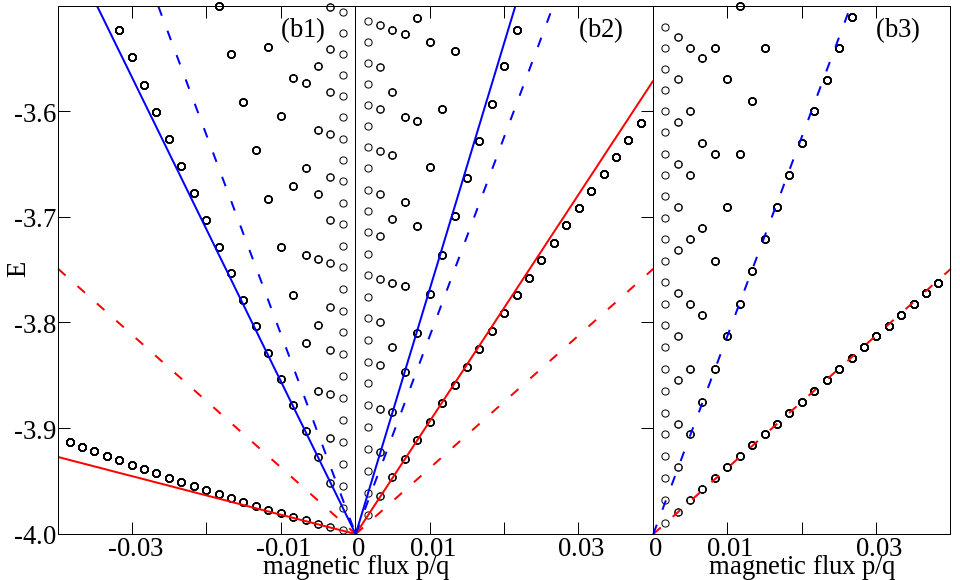}  %
\includegraphics[width=0.245\linewidth, angle=0]{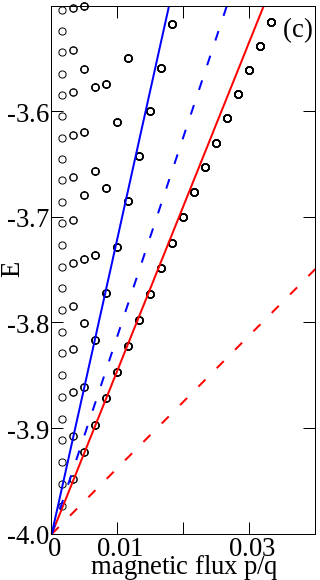}  %
\includegraphics[width=0.245\linewidth, angle=0]{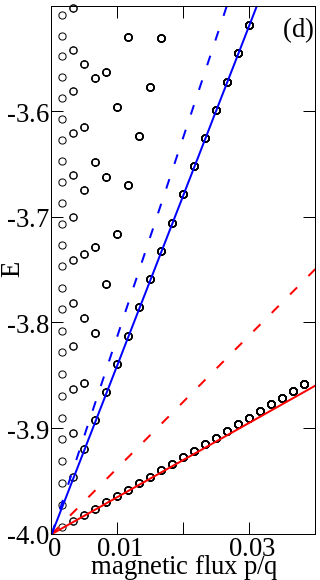}  %
\includegraphics[width=0.245\linewidth, angle=0]{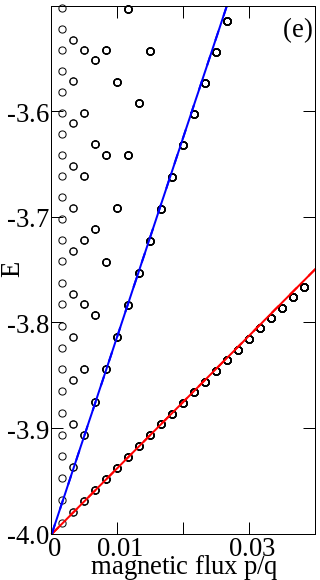}  %
\includegraphics[width=0.245\linewidth, angle=0]{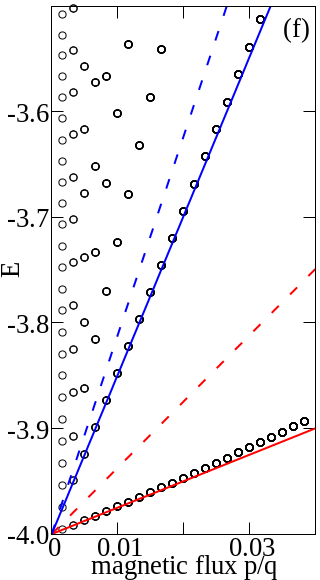}  %
\caption{(Color online.) Effect of circularly polarized light on the energy spectrum for Landau levels in a weak uniform magnetic flux $\phi$ on the square lattice. The circularly polarized light $\bfA(t) = A_0 (\sin\Omega t, \cos\Omega t)$ with frequency fixed at an off-resonant regime $\Omega = 8.0$. The representative laser laser intensity are chosen as (a) $A_0 = 0.0$, (b1-b3) $A_0 = 1.6$, (c) $A_0 = 2.0$ and (d) $A_0 = 3.0$, (e) $A_0 = 3.8$ and (f) $A_0 = 5.0$. The calculations are done numerically with 3 Floquet copies. The magnetic flux is defined as $\phi = p/q$ with $p$ ranging from 1 to $q-1$ and $q=599$.
The dashed lines are the semi-classical results for Landau levels with $E=\hbar\omega_c (n+1/2)$ where $n=1$ (red dashed lines) and $n=2$ (blue dashed lines). The red (blue) solid line is the semi-classical results plus the high-frequency expansion to 1st order of $\hbar\Omega$.
}
\label{fig:landau-offresonance}
\end{figure*}
\section{Effective Hamiltonian for small magnetic flux - Landau level regime}
\label{sec:landau}
In this section, we focus on the Landau level at small magnetic flux $\phi$.
Starting from the tight-bing Hamiltonian Eq.\eqref{eq:Hsqr} (absence of laser), for the general case $t_y = \lambda t_x=\lambda \tsh$ and small $k_x \ll 1, k_y \ll 1$, The band dispersion is written as,
\begin{equation}\begin{split}
	\epsilon_k =& -2\tsh (\cos(k_xa) + \lambda \cos(k_ya)) \nonumber\\
	\approx& -2\tsh (1-k_x^2a^2/2) - 2\lambda \tsh (1-k_y^2a^2/2) \nonumber\\
	      =& -2\tsh (1+\lambda) + (k_x^2 + \lambda k_y^2) \tsh a^2
\end{split}\end{equation}
which corresponds to the effective mass $m^* = \hbar^2/(2ta^2)$.
From the semi-classical theory,  we have the energy for system subjected to perpendicular uniform magnetic field\cite{Hatsuda:njp16} as,
\begin{equation}\begin{split}
	\epsilon(\phi) =& -2\tsh(1+\lambda) + \sqrt{\lambda} \hbar\omega_c (n+1/2) \\
                   =& -2\tsh(1+\lambda) + 4\pi \sqrt{\lambda} t\phi(n+1/2)
                   \label{eq:landau}
\end{split}\end{equation}
where we used $\omega_c = eB/m^* c$, and $n=0,1,2,\cdots$ is the serial number of Landau levels.
The landau levels are plotted in Fig.\ref{fig:landau-offresonance}(a) with the red and blue solid lines for Landau level with $n=0$ and $n=1$, respectively.

Let's go back to the deformed Hofsadter butterfly in off-resonance circularly polarized light.
Since the reflection symmetry about magnetic flux $\phi=1/2$ is preserved by the time-reversal symmetry and the circularly polarized light break that symmetry, which is $E(\phi) \neq E(-\phi) = E(1-\phi)$. To investigate
the influence of circularly polarized light on the Landau levels, we plot the energy as a function of $\phi$ from $-0.04<\phi<0.04$ in Fig.\ref{fig:landau-offresonance}(b1-b2). The dashed lines are the original Landau level without laser using Eq.\eqref{eq:landau}, the numerical results for $\phi>0$ is on top of the dashed line and $\phi<0$ is under the dashed line.
The averaged result $\epsilon_n'(\phi) = (\epsilon_n(\phi) + \epsilon_n(-\phi))/2$ is plotted as a function $\phi$ in Fig.\ref{fig:landau-offresonance}(b3). The averaged one for small flux is in good agreement with the semi-classical results in Eq.\eqref{eq:landau}.
Going back to the high frequency effective Hamiltonian. we have $H_{\text{eff}}(\phi) = H_0(\phi) + H'(\phi)$ with $H_0(\phi) = H_0(-\phi)$ and $H'(\phi) = -H'(-\phi)$. Here we calculate the commutator of $H_0(\phi)$ and $H'(\phi)$, we have,
\begin{equation}
    [H_0(\phi), H'(\phi)] = \mathcal{O}(\phi^2) \approx 0
    \label{H0Hpcom}
\end{equation}
which means the two operators have approximately the same eigenvectors and finally, the two Hamiltonian $H(\phi)$ and $H(-\phi)$ have the property $(E_{i}(\phi) + E_i(-\phi))/2 \approx -2t_x - 2t_y + 4\pi \lambda t_x \phi $ which is consistent with the semi-classical calculation.

From the semi-classical theory, we have the eigenvalues of $H_0$ is proportional to $\phi$. Here we realize that $H'(\phi) \propto \phi$ for small magnetic flux $\phi$, so we carefully check if the eigenvalues of the two are still proportional to $\phi$ ?
By diagonalizing the Hamiltonian $H_0$, which is
\begin{equation}
    H_0(\phi) |\Psi_i\rangle = E_i^0(\phi) |\Psi_i\rangle
\end{equation}
with $E_i^0(\phi)$ and $|\Psi_i\rangle$ the $i$-th eigenvalue and eigenvector. Note here we choose $\phi=1/q$ to aviod the degeneracy of the energy level because the eigenvectors can mix with each other for states with the same energy.
By considering the commutation relation $[H_0(\phi), H'(\phi)] \approx 0$ for small magnetic flux $\phi$, we have
\begin{equation}
    H'(\phi) |\Psi_i\rangle \approx E_i' |\Psi_i\rangle
\end{equation}
and the $i$-th eigenvalues of effective Hamitonian is $E_i + E_i'$.
we numerically find that
\begin{equation}
    E_{i}' \approx 6.25\times8\lambda t^2\mathcal{J}_1^2(A_0a)\phi/(\hbar\Omega)
\end{equation}
Combining with the result with laser (the Landau level), we have
\begin{equation}\begin{split}
	\epsilon(\phi)
                   =& (-2t-2\lambda t) + 4\pi \sqrt{\lambda} t\phi(n+1/2)\\
                   & + 6.25\times8\lambda t^2\mathcal{J}_1^2(A_0a)\phi/(\hbar\Omega)
                   \label{eq:mlandau}
\end{split}\end{equation}
where the co-efficient $6.25$ is determined numerically. The results are plotted as a solid red and blue lines in Fig.\ref{fig:landau-offresonance} with $n=0$ and $n=1$, respectively. The semi-classical results are in good agreement with the numerical data in the Landau regime. Our result show that, the influence of circularly laser on the Landau level is proportional to magnetic flux $\phi$ and independent of the serial number of Landau level.

\begin{figure*}[t]
\includegraphics[width=0.485\linewidth, angle=0]{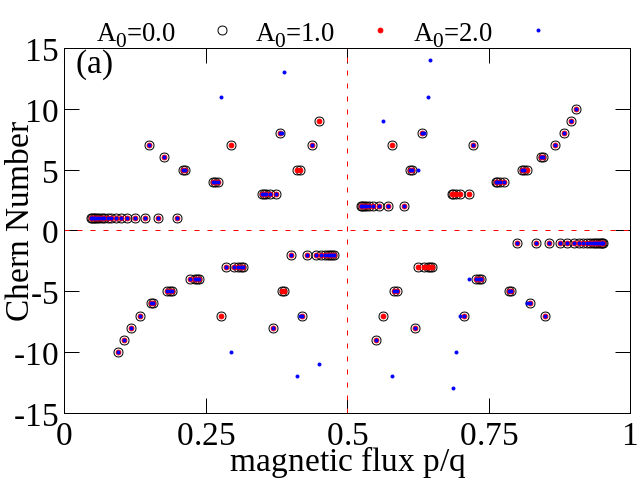}  %
\includegraphics[width=0.485\linewidth, angle=0]{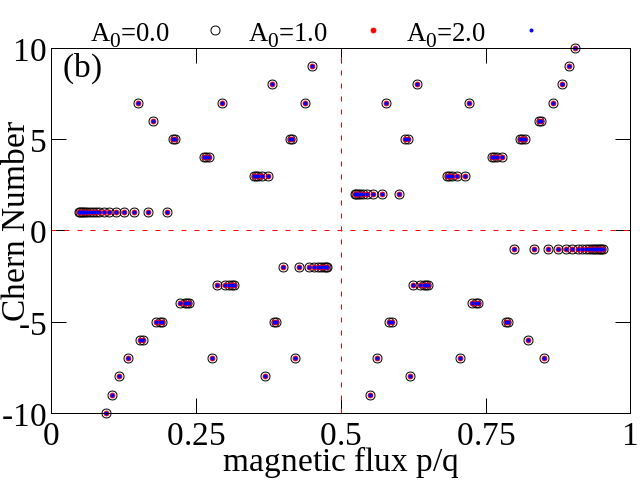}  %
\caption{(Color online) Chern number $|C_n|$ as a function of the magnetic flux $\Phi$ for the square lattice exposed to circularly (a) and linearly (b) polarized lasers with the laser frequency fixed at off-resonance region $\Omega = 8.0$. (a) circularly polarized light $A_0 = 1.0, 2.0$.
(b) linearly polarized light $A_0 = 1.0, 2.0, \alpha = \pi/6$. The calculations are done with 5 Floquet copies. The magnetic flux is defined as $\phi = p/q$ with $p$ ranging from 1 to $q-1$ and $q \leq 21$ and $(p,q)$ are co-prime number.}
\label{fig:chern}
\end{figure*}
\section{Spin Chern number for the driven Hofstadter butterfly}
\label{sec:chern}
Following the previous work by Goldman\cite{Goldman:jpb09} and Wackerl\cite{Wackerl:prb19}, we calculate the ground state Chern number of Hofstadter's butterfly, where the ``ground state'' in Floquet-Bloch band structure shall be understood as the lowest energy band of the central Floquet copy ($n-m=0$).
Here the stratege to calculate the Chern number is the Fukui's method\cite{Fukui:jpsj05} .
To clearly seperate bands with different Floquet copy index, we choose an off-resonance frequency $\hbar\Omega = 8.0$ (effective bandwidth is $W_{\mathrm{eff}} = 8 \mathcal{J}_0(A_0a)\le 8$).

In Fig.\ref{fig:chern}, we plot the Chern number for the ``ground state'' of the Hofstadter's butterfly with laser frequency fixed at $\hbar\Omega = 8.0$.
In all these Chern number plots, we plot the Chern number in equilibrium (absence of laser) as a reference point.
The time-reversal symmetry $\mathcal{T}H(\phi)\mathcal{T}^{-1} = H(-\phi) = H(1-\phi)$ preserve the symmetry of Chern number $C(\phi) = - C(1-\phi)$.

In Fig.\ref{fig:chern}(a), we plot the Chern number as a function of magnetic flux for circularly polarized light with parameter $A_0 = 1.0, 2.0$.
From the numerical data, we realize the Chern numbers at laser intensity $A_0 = 1.0$ are the same as the reference points. Further increasing the laser intensity to $A_0=2.0$ will show some difference, the Chern numbers differ somewhat from the reference points, especially at larger magnetic flux area (around $\phi=1/2$).

In Fig.\ref{fig:chern}(b), we plot Chern number as a function magnetic flux for linearly polarized light with laser intensity $A_0 = 1.0, 2.0$ and polarization direction along the $\alpha=\pi/6$. These data are the same as the one without drive. Further, we checked the Chern number for different polarization $\alpha=0, \pi/3, \pi/2$ and $A_0 = 1.0, 2.0$, the results are the same with the reference data. This can be understood using the effective Hamiltonian in Eq.\eqref{eq:hflin}, where $1/\hbar\Omega$ order terms vanish.

Since the magnetic-translation symmetry is preserved as the system is exposed to an external laser, the topological invariant must satisfy the Diophantine equation \cite{Lababidi:prl14,Kooi:prb18}
\begin{align}
   s = \frac{1}{q} + \frac{p}{q} C
   \label{eq:diophantine}
\end{align}
for flux $\phi = p/8q$, where $C$ is the topological invariant and $s$ is an integer. We have verified that our calculated Chern numbers satisfy the Diophantine equation.

\section{DISCUSSION AND CONCLUSIONS}
\label{sec:conclusion}
In this paper, we study the Hofstadter butterfly in the square lattice and the influence of a periodic driving laser (linearly or circularly polarized) on its fractal structure and its ``ground state'' Chern number using the Floquet theory. The energy spectrum in equilibrium (Hofstadter butterfly) exhibit self-similar fractal structure with rich symmetries. Qualitatively, we focus on three symmetries in this work, the particle-hole symmetry, the reflection symmetry about magnetic flux $\phi=1/2$, the inversion symmetry about $E=0, \phi=1/2$. The influence of circularly polarized light will in general break the time-reversal symmetry and sublattice symmetry, as a result, the particle-hole and reflection symmetry about $\phi=1/2$ are broken. By contrast, the inversion symmetry about $E=0, \phi=1/2$ is preserved, which is explained by considering the sublattice symmetry. Further increasing the laser intensity to critical value $A_0a=3.80$, the three symmetries are recovered approximately. This is explained using effective Hamiltonian derived through the Floquet-Magus expansion.
As the Hofstadter butterfly in square lattice is exposed by off-resonance linearly polarized light, the effective Hamiltonian is just the original one renomalized by $\mathcal{J}_0(A_0a\cos\alpha)$ in the $x$-direction and by $\mathcal{J}_0(A_0a\cos\alpha)$ in the $y$-direction.

By considering the spectrum in the Landau level regime, we derived the effective Hamiltonian at off-resonance regime, and numerically determined the influence of circularly polarized light. The Landau level spectrum is modified by,
$6.25\times8 t^2\mathcal{J}_1^2(A_0a)\phi/(\hbar\Omega)$ which depend on the laser intensity, frequency and in-dependent of Landau level serial number.

Finally, we calculate the ``ground state'' Chern number influenced by laser in off-resonance regime. For circularly polarized light, laser with small intensity will not change the Chern number. The one with large intensity will change Chern number, mainly at large magnetic field regime ($\phi=1/2$). The linear polarized light will not change the Chern number in our calculation for laser intensity $A_0=1.0,2.0$ and different polarization direction $\alpha =0, \pi/6, \pi/3, \pi/2$.
Our work highlights the generic features expected for the periodically driven Hofstadter problem on square lattice and provide the strategy to engineering the Hofstadter butterfly with laser.
\newpage
\newpage
\section{appendices}
\subsection{high frequency expansion}
Starting from Eq.\eqref{eq:hfcopy} and Eq.\eqref{eq:hfmaggcir}, we write $H^{n,m}$ or $H^{n-m}$ explicitly,
\begin{widetext}
\begin{equation}
	  H^{0}
	 = -t_x \mathcal{J}_{0}(A_0 a) \left(\sum_{l=1}^{q-1} a^{\dagger}_{l} a_{l+1}^{} + a^{\dagger}_{q} a_{1}^{}\right)
	   -t_x \mathcal{J}_{0}(A_0 a) \left(\sum_{l=1}^{q-1} a^{\dagger}_{l+1} a_{l}^{} + a^{\dagger}_{1} a_{q}^{}\right)
	   -2t_y \mathcal{J}_{0}(A_0a)   \sum_{l=1}^{q} a^{\dagger}_{l} a_{l}^{} \cos(2\pi l\phi)
\end{equation}
\begin{equation}
	  H^{1}
	 = +t_x \mathcal{J}_{1}(A_0 a) \left(\sum_{l=1}^{q-1} a^{\dagger}_{l} a_{l+1}^{}
        + a^{\dagger}_{q} a_{1}^{}\right)
	   -t_x \mathcal{J}_{1}(A_0 a) \left(\sum_{l=1}^{q-1} a^{\dagger}_{l+1} a_{l}^{}
        + a^{\dagger}_{1} a_{q}^{} \right)
	   + 2t_y \mathcal{J}_{1}(A_0a)\sum_{l=1}^{q} a^{\dagger}_{l} a_{l}^{} \sin(2\pi l\phi)
\end{equation}
\begin{equation}
	  H^{-1}
	 = - t_x \mathcal{J}_{1}(A_0a)\left(\sum_{l=1}^{q-1} a^{\dagger}_{l} a_{l+1}^{} +a^{\dagger}_{q} a_{1}^{} \right)
	   + t_x \mathcal{J}_{1}(A_0 a)\left(\sum_{l=1}^{q-1} a^{\dagger}_{l+1} a_{l}^{}  + a^{\dagger}_{q} a_{1}^{}\right)
	   +2 t_y \sum_{l=1}^{q} a^{\dagger}_{l} a_{l}^{} \mathcal{J}_{1}(A_0a) \sin[+ 2\pi l\phi]
\end{equation}
\begin{equation}
	H_{2} = H_{-2}
	 = -t_x \mathcal{J}_{2}(A_0 a) \left(\sum_{l=1}^{q-1} a^{\dagger}_{l} a_{l+1}^{} + a^{\dagger}_{q} a_{1}^{}\right)
	   -t_x \mathcal{J}_{2}(A_0 a) \left(\sum_{l=1}^{q-1} a^{\dagger}_{l+1} a_{l}^{} + a^{\dagger}_{1} a_{q}^{}\right)
	   - 2 t_y \mathcal{J}_{2}(A_0a) \sum_{l=1}^{q} a^{\dagger}_{l} a_{l}^{} \cos[2\pi l\phi]
\end{equation}
\begin{equation}
	  H_{3}
	 = +t_x \mathcal{J}_{3}(A_0 a) \left(\sum_{l=1}^{q-1} a^{\dagger}_{l} a_{l+1}^{}  + a^{\dagger}_{q} a_{1}^{} \right)
	   -t_x \mathcal{J}_{3}(A_0 a) \left(\sum_{l=1}^{q-1} a^{\dagger}_{l+1} a_{l}^{} + a^{\dagger}_{1} a_{q}^{}\right)
	   -2 t_y \mathcal{J}_{3}(A_0a) \sum_{l=1}^{q} a^{\dagger}_{l} a_{l}^{} \sin[+ 2\pi l\phi]
\end{equation}
\begin{equation}
	  H_{-3}
	 = - t_x \mathcal{J}_{3}(A_0 a) \left(\sum_{l=1}^{q-1} a^{\dagger}_{l} a_{l+1}^{}  + a^{\dagger}_{q} a_{1}^{} \right)
	   + t_x \mathcal{J}_{3}(A_0 a) \left(\sum_{l=1}^{q-1} a^{\dagger}_{l+1} a_{l}^{}  + a^{\dagger}_{q} a_{1}^{} \right)
	   -2t_y \mathcal{J}_{3}(A_0a) \sum_{l=1}^{q} a^{\dagger}_{l} a_{l}^{} \sin(2\pi l\phi)
\end{equation}
\end{widetext}
we have
\begin{widetext}
\begin{equation}\begin{split}
	[H_1,H_{-1}]
=8 t_xt_y\mathcal{J}_{1}^2(A_0 a)\sin(\pi \phi)
	\left(\sum_{l=1}^{q-1}a^{\dagger}_{\bfk,l} a_{\bfk,l+1}^{}\cos(2\pi (l+1/2)\phi)
	+ a^{\dagger}_{\bfk,q} a_{\bfk,1}^{}\cos(+ \pi \phi)\right) + h.c.
\end{split}\end{equation}

\begin{equation}\begin{split}
	[H_3,H_{-3}]
     &= -8 t_xt_y\mathcal{J}_{3}^2(A_0 a)\sin(\pi \phi)
	\left(\sum_{l=1}^{q-1}a^{\dagger}_{\bfk,l} a_{\bfk,l+1}^{}\cos(2\pi (l+1/2)\phi)
	+ a^{\dagger}_{\bfk,q} a_{\bfk,1}^{}\cos(+ \pi \phi)\right) + h.c.
\end{split}\end{equation}
\end{widetext}
\subsection{Proof of $[H_0, [H_1, H_{-1}]] \propto \phi^2$ for $\phi\ll 1$}
By setting,
\begin{equation}\begin{split}
      \alpha_0 = -t_x \mathcal{J}_{0}(A_0 a),\quad
      \beta_0  = -2 t_y \mathcal{J}_{0}(A_0 a) \nonumber\\
      \alpha_1 = -t_x \mathcal{J}_{1}(A_0 a), \quad
      \beta_1  = -2 t_y \mathcal{J}_{1}(A_0 a), \nonumber\\
\end{split}\end{equation}
\begin{equation}\begin{split}
	 X = \sum_{l=1}^{q-1} a^{\dagger}_{\bfk,l} a_{\bfk,l+1}^{} + a^{\dagger}_{\bfk,q} a_{\bfk,1}^{},
\end{split}\end{equation}
\begin{equation}\begin{split}
	 Y = \sum_{l=1}^{q} a^{\dagger}_{\bfk,l} a_{\bfk,l}^{} \cos(2\pi l\phi),
\end{split}\end{equation}
\begin{equation}\begin{split}
     Z = \sum_{l=1}^{q} a^{\dagger}_{\bfk,l} a_{\bfk,l}^{} \sin(2\pi l\phi),
\end{split}\end{equation}
we have,
\begin{equation}\begin{split}
	  H_{+0} = +\alpha_0 (X+X^\dagger) + \beta_0 Y, \nonumber\\
	  H_{+1} = -\alpha_1 (X-X^\dagger) - \beta_1 Z, \nonumber\\
	  H_{-1} = +\alpha_1 (X-X^\dagger) - \beta_1 Z.
\end{split}\end{equation}
The commutator between $H_1$ and $H_{-1}$ is expressed as,
\begin{equation}
[H_1, H_{-1}] = 2\alpha_1\beta_1 [X, Z] + 2 \alpha_1\beta_1 [X, Z]^\dagger.
\end{equation}
Further, the commutator between $H_0$ and $[H_1, H_{-1}]$ is expressed as,
\begin{equation}\begin{split}
&[H_0, [H_1, H_{-1}]] \\
                     &= 2\alpha_0\alpha_1\beta_1 \left([X, [X, Z]]  - [X, [X, Z]]^\dagger\right) \\
                     &+ 2\alpha_1\beta_0 \beta_1 \left([Y, [X, Z]]  - [Y, [X, Z]]^\dagger\right) \\
                     &+ 2\alpha_0\alpha_1\beta_1 \left([X, [X, Z]^\dagger]  - [X, [X, Z]^\dagger]^\dagger\right)
\end{split}\end{equation}

\begin{widetext}
\begin{equation}\begin{split}
	  [X, Z]
                     &= 2 \sin(\pi\phi) \left( \sum_{l=1}^{q-1} a^{\dagger}_{l} a_{l+1}^{} \cos(2\pi (l+1/2)\phi)+  a^{\dagger}_{q} a_{1}^{} \cos(\pi\phi)\right)
\end{split}\end{equation}

\begin{equation}\begin{split}
	  [X,[X, Z]]
=& -4\sin^2(\pi \phi)\left(\sum_{l=1}^{q-2} a^{\dagger}_{l} a_{l+2}^{} \sin(2\pi (l+1)\phi)
    +a^{\dagger}_{q} a_{2}^{}\sin(2\pi\phi)\right)
\end{split}\end{equation}
\begin{equation}\begin{split}
	  [X,[X, Z]^\dagger]
=&-4 \sin^2(\pi \phi)\left(\sum_{l=2}^{q-1} a^{\dagger}_{l} a_{l}^{}\sin(2\pi l\phi)
 + a^{\dagger}_{1} a_{1}^{} \sin(2\pi \phi) \right)
\end{split}\end{equation}

\begin{equation}\begin{split}
 [Y,[X, Z]]
 =& 2 \sin^2(\pi \phi) \left( \sum_{l=1}^{q-1} a^{\dagger}_{l} a_{l+1}^{} \sin(2\pi (2l+1)\phi)
     + a^{\dagger}_{q} a_{1}^{}\sin(2\pi\phi) \right)
\end{split}\end{equation}
\end{widetext}
As a result, we have $[H_0, [H_1, H_{-1}]] \propto \phi^2$ for $\phi\ll 1$, and operator $H_0$ have approximately the same eigenvectors with  $H'=[H_1, H_{-1}]$.

\subsection{Inversion symmetry about $E=0, \phi=1/2$}
\begin{figure}[t]
\includegraphics[width=1.0\linewidth]{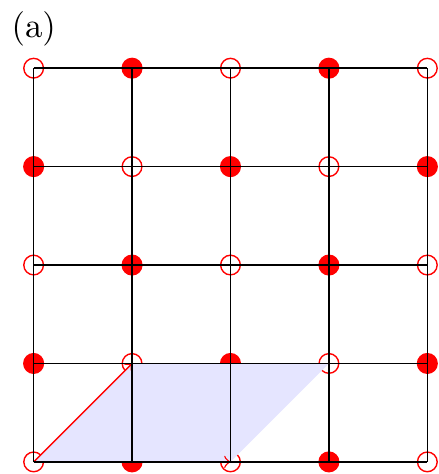}  %
\caption{(Color online) (a) Square lattice with A-B sub-lattice in one unit cell. The translational vectors are $\bfa_1 =(2,0)a, \bfa_2 = (1,1)a$ with $a$ lattice constant.}
\label{fig:square-sublat}
\end{figure}
Let's consider the A-B sub-lattice in the square lattice in Fig.\ref{fig:square-sublat}.
The model Hamiltonian is defined on the two dimensional square lattice with nearest-neighbour hopping terms,
\begin{equation}\begin{split}
	H = &\sum_{ij} - t_x\left( a^{\dagger}_{i,j} b_{i,j  } + a^{\dagger}_{i,j} b_{i-1,j  } + h.c. \right) \nonumber\\
	  + &\sum_{ij} - t_y\left( a^{\dagger}_{i,j} b_{i,j-1} + a^{\dagger}_{i,j} b_{i-1,j+1}  + h.c. \right)
\label{Habsqr}
\end{split}\end{equation}
where $t_x$ ($t_y$) is the hopping integral between nearest-neighbours along the $x$ ($y$) directions, $a^{\dagger}_{i,j} (b_{i,j})$ create (annihilate) an electron (the spin index is omitted for simplicity)
at sub-lattice A (B) in the unit $(i,j)$.

The position of an arbitrary unit cell is
\begin{equation}
    \bfR(i,j) = i \bfa_{1} + j\bfa_{2}
\end{equation}
where $i,j$ are integers, $\bfa_1=(2,0)a$ and $\bfa_2=(1,1)a$ are the translational lattice vectors with $a$ the lattice constant.

As the system is exposed to field (electric, magnetic field) with vector potential $\bfA=(A_x, A_y)$,
the hopping terms are modified by the Peierls substitution,
\begin{equation*}
    a_{i,j}^\dagger b_{i',j'}\mapsto \exp\left(+i\frac{e}{\hbar c}\int_{{R_b}(i',j')}^{{R_a}(i,j)}
    \bfA\cdot d\bfr\right) a_{i,j}^\dagger b_{i',j'}
\label{PeierlsPhase}
\end{equation*}
where $R_a(i,j)$ is the coordinate of sub-lattice A with unit cell index $(i,j)$, with $R_a(i,j)=m \bfa_{1} + n\bfa_{2}$ and $R_b(i,j)=i \bfa_{1} + j\bfa_{2} + (1,0)a$.
\subsubsection{Model Hamiltonian exposed to laser}
As the system is exposed to laser with vector potential $A_l(t) = [A_{lx}(t), A_{ly}(t)]$,
the Hamiltonian in real space is modified through the Perils substitution,
\begin{equation}\begin{split}
	H = &\sum_{ij} - t_x\left( a^{\dagger}_{i,j} b_{i,j  } \exp(-i A_{lx}(t)a) + h.c. \right) \nonumber\\
	  + &\sum_{ij} - t_x\left( a^{\dagger}_{i,j} b_{i-1,j} \exp(+i A_{lx}(t)a) + h.c.\right) \nonumber\\
	  + &\sum_{ij} - t_y\left( a^{\dagger}_{i,j} b_{i-1,j+1}\exp(-i A_{ly}(t)a)  + h.c. \right)\nonumber\\
	  + &\sum_{ij} - t_y\left( a^{\dagger}_{i,j} b_{i,j-1}\exp(+i A_{ly}(t)a) + h.c. \right)
\label{Habsqr}
\end{split}\end{equation}
For circularly polarized light with vector potential $A_{l}(t) = A_0 (\sin\Omega t, \cos\Omega t)$, we have
\begin{equation}\begin{split}
	&H^{nm} = - t_x\sum_{ij} \left( a^{\dagger}_{i,j} b_{i,j  } + b^{\dagger}_{i-1,j} a_{i,j} \right) \mathcal{J}_{m-n}(+A_0 a) \nonumber\\
	  &- t_x\sum_{ij} \left(b^{\dagger}_{i,j} a_{i,j  } + a^{\dagger}_{i,j} b_{i-1,j} \right)\mathcal{J}_{m-n}(-A_0 a) \nonumber\\
	  &- t_y \sum_{ij} \left( a^{\dagger}_{i,j} b_{i-1,j+1} + b^\dagger_{i,j-1}a_{i,j} \right)i^{n-m} \mathcal{J}_{m-n}(+A_0 a) \nonumber\\
	  &- t_y \sum_{ij} \left( b^\dagger_{i-1,j+1} a_{i,j} + a^{\dagger}_{i,j} b_{i,j-1} \right) i^{n-m}\mathcal{J}_{m-n}(-A_0 a)
\label{Habsqr}
\end{split}\end{equation}
\subsubsection{Hamiltonian with magnetic field}
As the system is exposed to perpendicular magnetic field $\bfB =\nabla\times \bfA_m= (0,0, -B)$ and
adopt the Landau gauge with vector potential $\bfA_m = (By, 0, 0)$.
As usual, we restrict the  flux per unit cell
in units of the elementary charge over Planck's constant to a rational value
\begin{equation}
    \phi \equiv B a^2.
\end{equation}
Thus, the Peierls phase can be written as
\begin{equation}
    \frac{e}{\hbar c} B a^2 = 2\pi \phi/\phi_0
\label{Peierls}
\end{equation}
where $\phi_0 = hc/e$ is the magnetic quantum flux, $\phi=p/q$ with $(p,q)$ co-prime integers.
To recover the translational symmetry of the lattice, we enlarge the unit cell along the translational vector $\bfa_1$ by $q$ time of the original unit cell without magnetic field. The Hamiltonian in real-space can be rewritten as,
\begin{equation}\begin{split}
 H = & -t_x \sum_{ij}\sum_{l=1}^{q} a^{\dagger}_{i,j,l} b_{i,j,l}^{} e^{-i2\pi l\phi} + h.c.\nonumber\\
     & -t_x \sum_{ij}\sum_{l=1}^{q} a^{\dagger}_{i,j,l} b_{i-1,j,l}^{} e^{+i2\pi l\phi} + h.c.\nonumber\\
     & -t_y \sum_{ij}\sum_{l=2}^{q} a^{\dagger}_{i,j,l} b_{i,j,l-1}^{} +a^{\dagger}_{i,j,1} b_{i,j-1,q}^{} + h.c.\nonumber\\
     & -t_y \sum_{ij}\sum_{l=1}^{q-1} a^{\dagger}_{i,j,l} b_{i-1,j,l+1}^{}  +a^{\dagger}_{i,j,q} b_{i-1,j+1,1} + h.c.
\end{split}\end{equation}
where the position of unit cell is
\begin{equation}
\tilde{\bf{R}}(i,j) = i \bfa_{1}  + j\bfa_{2} \times q
\end{equation}
After Fourier transformation,
\begin{equation}\begin{split}
	H_{\bfk}
	 =&-t_x \sum_{l=1}^{q} a^{\dagger}_{\bfk,l} b_{\bfk,l}^{} \left( e^{-i 2\pi l \phi}
	  + e^{i \bfk \cdot \bfa_1} e^{+i 2\pi l \phi}\right) + h.c. \nonumber\\
	 &-t_y \sum_{l=2}^{q} a^{\dagger}_{\bfk,l} b_{\bfk,l-1}^{} + a^{\dagger}_{\bfk,q} b_{\bfk,1}^{} e^{i \bfk \cdot q\bfa_2} + h.c. \nonumber\\
	 &-t_y \sum_{l=1}^{q-1} a^{\dagger}_{\bfk,l} b_{\bfk,l-1}^{} e^{i \bfk \cdot \bfa_1} + a^{\dagger}_{\bfk,1} b_{\bfk,q}^{} e^{i \bfk \cdot (\bfa_1-q\bfa_2)} + h.c.
\end{split}\end{equation}
For the case $\bfk=(0,0)$ (subscript $\bfk$ is ignored for simplicity), we have
\begin{equation}\begin{split}
	H_{\mathrm{butterfly}}
	 =&-t_x \sum_{l=1}^{q} a^{\dagger}_{l} b_{l}^{} \left( e^{-i 2\pi l \phi}
	  + e^{+i 2\pi l \phi}\right) + h.c. \nonumber\\
	 &-t_y \sum_{l=2}^{q} a^{\dagger}_{l} b_{l-1}^{} + a^{\dagger}_{q} b_{1}^{}  + h.c. \nonumber\\
	 &-t_y \sum_{l=1}^{q-1} a^{\dagger}_{l} b_{l-1}^{}  + a^{\dagger}_{1} b_{q}^{} + h.c.
	 \label{eq:eqbtfy}
\end{split}\end{equation}
This is the original Hamiltonian to generate the Hofstadter butterfly on the square lattice, with A-B sublattice considered.

The matrix form Hamiltonian is expressed as,
\begin{equation}
    \begin{split}
        H = \begin{pmatrix}
      0 & H_{\mathrm{AB}} \\
      H_{\mathrm{AB}}^\dagger & 0\\
      \end{pmatrix}
    \end{split}
\end{equation}
\begin{widetext}\begin{equation*}
H_{\mathrm{AB}} =
\begin{pmatrix}
      a^{\dagger}_{1} \\
      a^{\dagger}_{2} \\
      \vdots \\
      a^{\dagger}_{q} \\
\end{pmatrix}^{\mathrm{T}}
\begin{pmatrix}
	-2t_x\cos(2\pi\phi) & -t_y & \cdots & -t_y  \\
         -t_y & -2t_x\cos(4\pi\phi) & \cdots & 0 \\
         \vdots  & \vdots  & \ddots & \vdots  \\
         -t_y & 0 & \cdots & -2t_x\cos(2\pi q\phi)
\end{pmatrix}
\begin{pmatrix}
	b_{1}^{} \\
	b_{2}^{} \\
        \vdots \\
	b_{q}^{} \\
\end{pmatrix}
\end{equation*}\end{widetext}
Apparently, the A-B sublattice symmetry in square lattice will preserve the particle-hole symmetry in the original Hofstadter butterfly\cite{Das:prb2020}.

\subsubsection{Hamiltonian with both laser and magnetic field}
As the system is exposed to laser, the Hamiltonian is rewritten as,
The Hamiltonian in real-space is,
\begin{widetext}
\begin{equation}\begin{split}
 &H =  -t_x \sum_{ij}\sum_{l=1}^{q} a^{\dagger}_{i,j,l} b_{i,j,l}^{} e^{-i2\pi l\phi} \exp(-i A_{lx}(t)a) + h.c.\nonumber\\
     & -t_x \sum_{ij}\sum_{l=1}^{q} a^{\dagger}_{i,j,l} b_{i-1,j,l}^{} e^{+i2\pi l\phi} e^{+i A_{lx}(t)a} + h.c.\nonumber\\
     & -t_y \sum_{ij}\left(\sum_{l=2}^{q} a^{\dagger}_{i,j,l} b_{i,j,l-1}^{} +a^{\dagger}_{i,j,1} b_{i,j-1,q}^{} \right) e^{-i A_{ly}(t)a} + h.c.\nonumber\\
     & -t_y \sum_{ij}\left(\sum_{l=1}^{q-1} a^{\dagger}_{i,j,l} b_{i-1,j,l+1}^{}  +a^{\dagger}_{i,j,q} b_{i-1,j+1,1}\right) e^{+i A_{ly}(t)a} + h.c.
\end{split}\end{equation}

Fourier transformation and set $\bfk=(0,0)$,
\begin{equation}\begin{split}
	H_{\bfk}
	 =&-t_x \left(\sum_{l=1}^{q} a^{\dagger}_{l} b_{l}^{} e^{-i 2\pi l \phi} e^{-i A_{lx}a}
	  + a^{\dagger}_{l} b_{l}^{} e^{+i 2\pi l \phi} e^{+i A_{lx}(t)a} \right) + h.c. \nonumber\\
	 &-t_y \sum_{l=1}^{q} a^{\dagger}_{l} b_{l-1}^{} e^{-i \bfA_{ly}(t)a} + a^{\dagger}_{1} b_{q}^{} e^{-i A_{ly}(t)a}  + h.c. \nonumber\\
	 &-t_y \sum_{l=1}^{q} a^{\dagger}_{l} b_{l+1}^{} e^{+i A_{ly}(t)a}  + a^{\dagger}_{q} b_{1}^{} e^{+i A_{ly}(t)a} + h.c.
	 \label{eq:eqbtfy}
\end{split}\end{equation}
\end{widetext}
\subsubsection{Circularly polarized light}
For circularly polarized light with vector potential $\bfA(t)=A_0[\sin(\Omega t), \cos(\Omega t)]$, the matrix elements of the Floquet-Bloch Hamiltonian are
\begin{equation}
     H_{nm} = \frac{1}{T} \int_0^T dt e^{-i(n-m)\Omega t}  H(t)
\end{equation}
we need to calculate the expression with the general form as,
\begin{equation}
      f_{nm} = \frac{1}{T} \int_0^T dt e^{-i(n-m)\Omega t}  \exp[-i \bfA(t)\cdot{\bf d}]
\end{equation}
Here we used ${\bf d}={\bf R}_j-{\bf R}_i$.
Substituting the vector potential of circular polarized light into above equation,
For ${\bf d} = (a, 0)$,
\begin{equation*} \begin{split}
	\frac{1}{T} \int_0^T dt e^{-i(n-m)\Omega t}  \exp[-i A_0 a \sin(\Omega t)] = \mathcal{J}_{m-n}(A_0 a)
\end{split}\end{equation*}
For ${\bf d} = (0, a)$,
\begin{equation*} \begin{split}
	&\frac{1}{T} \int_0^T dt e^{-i(n-m)\Omega t}  \exp[-i A_0 a \cos(\Omega t)] \nonumber\\
      = &\mathcal{J}_{m-n}(A_0a) \exp[i(n-m)\pi/2]
\end{split}\end{equation*}
where $\mathcal{J}_{n}(x)$ is the Bessel function of first kind.
The Floquet-Bloch Hamiltonian at  $\bfk=(0,0)$ for circularly polarized light is,
\begin{widetext}
\begin{equation}\begin{split}
	H_{\mathrm{btfy}}^{nm}(\phi)
	 =&-t_x \left(\sum_{l=1}^{q} \mathcal{J}_{m-n}(+A_0 a) a^{\dagger}_{l} b_{l}^{} e^{-i 2\pi l \phi}
	  + \mathcal{J}_{m-n}(-A_0 a)b^{\dagger}_{l} a_{l}^{} e^{+i 2\pi l \phi}\right) \nonumber\\
     &-t_x \left(\sum_{l=1}^{q} \mathcal{J}_{m-n}(-A_0 a) a^{\dagger}_{l} b_{l}^{} e^{+i 2\pi l \phi}
	  + \mathcal{J}_{m-n}(+A_0 a)b^{\dagger}_{l} a_{l}^{} e^{-i 2\pi l \phi}\right) \nonumber\\
	 &-t_y \sum_{l=1}^{q-1} \mathcal{J}_{m-n}(-A_0 a) a^{\dagger}_{l} b_{l+1}^{} e^{i(n-m)\pi/2}+ \mathcal{J}_{m-n}(-A_0 a)a^{\dagger}_{q} b_{1}^{} e^{i(n-m)\pi/2}\nonumber\\
     &-t_y \sum_{l=1}^{q-1} \mathcal{J}_{m-n}(+A_0 a) b_{l+1}^{} a^{\dagger}_{l} e^{i(n-m)\pi/2}+ \mathcal{J}_{m-n}(+A_0 a)b^{\dagger}_{1} a_{q}^{} e^{i(n-m)\pi/2} \nonumber\\
	 &-t_y \sum_{l=2}^{q} \mathcal{J}_{m-n}(+A_0 a)a^{\dagger}_{l} b_{l-1}^{} e^{i(n-m)\pi/2} + \mathcal{J}_{m-n}(+A_0 a)a^{\dagger}_{1} b_{q}^{} e^{i(n-m)\pi/2}\nonumber\\
     &-t_y \sum_{l=2}^{q} \mathcal{J}_{m-n}(-A_0 a) b^{\dagger}_{l-1} a_{l} e^{i(n-m)\pi/2} + \mathcal{J}_{m-n}(-A_0 a)b^{\dagger}_{q} a_{1}^{} e^{i(n-m)\pi/2} = H_{\mathrm{btfy}}^{mn}(-\phi)
	 \label{eq:eqbtfy}
\end{split}\end{equation}
\end{widetext}
By considering the sublattice symmetry of square lattice,
we introduce a diagonal matrix $\sigma_z$ that equals $+1$ for sites on sub-lattice  A,
and $-1$ for sites on sub-lattice B, we can write the sub-lattice operation of the Hamiltonian as
\begin{equation}
   \sigma_z H_{\mathrm{btfy}}^{nm}(\phi) \sigma_z = \sigma_z H_{\mathrm{btfy}}^{mn}(-\phi) \sigma_z =  - H_{\mathrm{btfy}}^{mn}(-\phi)
\end{equation}
\begin{widetext}\begin{equation*} \begin{split}
H_{\mathrm{Floquet}}(\phi) =
&\begin{pmatrix}
    \ddots & \vdots & \vdots & \vdots & \udots \\
	\cdots & H_0(\phi) - \hbar\Omega \mathbb{1} & H_{-1}(\phi) & H_{-2}(\phi) & \cdots \\
    \cdots & H_1(\phi) & H_0(\phi) & H_{-1}(\phi) & \cdots \\
    \cdots & H_2(\phi) & H_1(\phi) & H_0(\phi) + \hbar\Omega \mathbb{1} & \cdots \\
    \udots & \vdots & \vdots & \vdots & \ddots
\end{pmatrix}\\
\sigma_z H_{\mathrm{Floquet}}(\phi) \sigma_z = -
&\begin{pmatrix}
    \ddots & \vdots & \vdots & \vdots & \udots \\
	\cdots & H_0(-\phi) + \hbar\Omega \mathbb{1} & H_{-1}(-\phi) & H_{-2}(-\phi) & \cdots \\
    \cdots & H_1(-\phi) & H_0(\phi) & H_{-1}(-\phi) & \cdots \\
    \cdots & H_2(-\phi) & H_1(\phi) & H_0(-\phi) - \hbar\Omega \mathbb{1} & \cdots \\
    \udots & \vdots & \vdots & \vdots & \ddots
\end{pmatrix}
= - H_{\mathrm{Floquet}}(-\phi)
\end{split}\end{equation*}\end{widetext}
As a result of the sub-lattice symmetry operation above, we proved that the eigenvalues of $H_{\mathrm{Floquet}}(\phi)$
is the inverse of the eigenvalues of $H_{\mathrm{Floquet}}(-\phi)$, which explains the inversion symmetry about $E=0,\phi=1/2$.
\acknowledgements
We acknowledge helpful discussions with Xiaoting Zhou. We gratefully acknowledge funding from the National Natural Science Foundation of China (Grant No. 11764017, 11904143, 11547184), Natural Science Foundation of Guangxi Province (Grant No. 2018GXNSFBA281003, No.2019GXNSFAA245034, No. AD19245180),
Foundation of promoting basic scientific research ability for
young teachers in universities of Guangxi (Grant No. 2019KY0086).
\bibliography{squarebutterfly}
\end{document}